\documentclass[preprint,showpacs,preprintnumbers,amsmath,amssymb,aps,pre]{revtex4}

\usepackage{graphicx}% Include figure files
\usepackage{dcolumn}% Align table columns on decimal point
\usepackage{bm}% bold math
\begin{document}

\preprint{LA-UR 05-8996}

\title{Time correlation functions in Vibration-Transit theory of liquid dynamics}

\author{Eric D. Chisolm}
\author{Giulia De Lorenzi-Venneri}
\author{Duane C. Wallace}

\affiliation{Theoretical Division, Los Alamos National Laboratory, 
Los Alamos, New Mexico 87545}

\date{\today}

\begin{abstract}
Within the framework of V-T theory of monatomic liquid dynamics, an
exact equation is derived for a general equilibrium time correlation
function. The purely vibrational contribution to such a 
function expresses the system's motion in one extended harmonic random
valley.  This contribution is analytically tractable and has no adjustable
parameters. While this contribution alone dominates the thermodynamic
properties, both vibrations and transits will make important
contributions to time correlation functions. By way of example, the
V-T formulation of time correlation functions is applied to the
dynamic structure factor $S(q,\omega)$.  The vibrational contribution
alone is shown to be in near perfect agreement with low-temperature
molecular dynamics simulations, and a model simulating the transit
contribution with three adjustable parameters achieves equally good
agreement with molecular dynamics results in the liquid regime.  The
theory indicates that transits will broaden without shifting the
Rayleigh and Brillouin peaks in $S(q,\omega)$, and this behavior is
confirmed by the MD calculations. We find the vibrational contribution
alone gives the location and much of the width of the liquid-state
Brillouin peak. We also discuss this approach to liquid dynamics compared 
with potential energy landscape formalisms and mode coupling  theory,
drawing attention to the distinctive features of our approach and to some 
potential energy landscape results which support our picture of the liquid state.

\end{abstract}

\pacs{05.20.Jj, 63.50.+x, 61.20.Lc, 61.12.Bt}% PACS, the Physics and Astronomy
                             % Classification Scheme.
\keywords{Liquid Dynamics, Inelastic Neutron Scattering, Intermediate scattering function}
\maketitle

\section{Introduction}

Long ago, Frenkel \cite{E1,E2} noted that the liquid-solid phase transition has
only a small effect on volume, cohesive forces, and specific heat, while the liquid
diffuses much more rapidly than the solid. From these facts he argued that the
motion of a liquid atom consists of approximately harmonic oscillations about an
equilibrium point, while the equilibrium point itself jumps from time to time.
Following this picture, Singwi et al. \cite{E3,E4a,E4b,E4c,E5} studied a series of models in which
a molecule undergoes vibrational motion for a time, then undergoes continuous
diffusion for a time. With adjustable parameters, the model was able to account
for incoherent parts of the neutron scattering data for water and lead. In discussing 
the theory of supercooled liquids and the glass transition, Goldstein \cite{E6}
presented an insightful description  of the potential energy landscape, and of the
vibrations about local minima and the transitions over barriers. From computer
simulations, Stillinger and Weber \cite{E11,E12,E13} found the local minima and named
them ``inherent structures". They suggested that the equilibrium properties of liquids
result from vibrational excitations within, and shifting equilibria between, these
structures. In the same spirit, Zwanzig \cite{E8} suggested an approximation for the
velocity autocorrelation function, in which that function evaluated for the vibrations
within a representative potential energy valley is multiplied by a relaxation
factor to account for the hops between valleys. Following these ideas, an active
research program has developed, which will be summarized in Sec.\;V.

In developing V-T theory for monatomic liquids, our initial goal was to construct 
an approximate but reasonably accurate Hamiltonian, from which the
partition function can be calculated analytically and without adjustable 
parameters \cite{VT5}. This is the most basic formulation available to condensed matter theory, and 
it was not available for liquids prior to V-T theory. In constructing a tractable model for the potential
energy surface, the following three new results were established.
(1) From extensive and highly accurate analysis of experimental specific heat data for
the elemental liquids at melt, the atomic motion was shown to consist almost
entirely of vibrations within nearly-harmonic many-atom valleys \cite{VT5}.
(2) From highly accurate analysis of experimental entropies of melting of the elements,
the disordering entropy was shown to be a universal constant plus
small scatter \cite{VT2,VT3,Vt4}. (3) From symmetry considerations it was concluded that
 the noncrystalline valleys
belong to two classes \cite{VT5}: (a) symmetric valleys, which have remnant crystalline
symmetry and consequently have a broad range of structural potentials and
vibrational frequency distributions, and (b) random valleys, which are of
overwhelming numerical superiority and which all have the same structural
potential and vibrational frequency distribution. This picture of the potential
surface has since been verified by computer studies \cite{VT8,VT9}. Only the
random valleys need to be considered further, since they completely dominate
the potential surface.  The number of random valleys is fixed by the universal disordering
entropy of the elemental liquids \cite{VT5}. 
Defining an extended random 
valley to be the harmonic extension to infinity of a random valley, with 
intervalley intersections ignored,
the leading order Hamiltonian is the sum of
extended random valley Hamiltonians, the corresponding partition function is called the
vibrational partition function, and any statistical 
mechanical quantity calculated using this Hamiltonian is called its vibrational
contribution \cite{VT5}. 
The vibrational contribution alone gives an accurate account of the thermodynamic
properties of the elemental liquids, without adjustable parameters \cite{VT5, VT6,VT14}. 
No other theory does this. Beyond the vibrational contribution, 
 corrections for anharmonicity and for the presence of intervalley intersections are well
defined and, though complicated, they are small \cite{VT5}.

In V-T theory, the motion of a liquid system across the boundary between
two random valleys is a transit. We have argued that, because of the
local character of equilibrium fluctuations in a many particle system, each
transit is accomplished by a small local group of atoms \cite{VT5}. Further, because
of the dominance of the vibrational contribution to thermodynamic functions,
the intervalley intersections must be rather sharp, and transits nearly
instantaneous \cite{VT5}. These predictions have since been verified by computer
simulations \cite{VT12}. In equilibrium, transits are crucial in allowing the system
to visit all the random valleys and thus exhibit the correct liquid entropy.
Transits are the expression in V-T theory of the equilibrium jumps of 
Frenkel \cite{E2,E1}, and of the
barrier hops of Goldstein \cite{E6} and Stillinger and Weber \cite{E11,E12,E13}.

In practical applications, V-T theory remains accurate when anharmonicity is
neglected entirely. Then the motion consists of the vibrational contribution, for
which we have exact analytic equations, plus transits over perfectly sharp
intervalley intersections, whose effect we have modeled. In this way, a
one-parameter model accounts for the temperature dependence of the specific heat
of liquids, as exemplified by the experimental data for mercury \cite{VT6}. For a model
system, equations were written for the liquid free energy, the glass free energy,
and for the nonequilibrium rate processes which occur in the liquid-glass
transition region where the free energy is not defined \cite{VT10}. This model
qualitatively reproduces the results of massive rate-dependent MD calculations
of Vollmayr, Kob, and Binder \cite{VKBJCP105}. A one-parameter model for the velocity
autocorrelation function gives a good account of MD  calculations for temperatures
from zero to $3T_{m}$ \cite{VT11}. Hence we have verified the applicability
of the equations of V-T theory for the glass, for the glass-transition
region, and for the liquid to very high temperatures. A review of V-T theory and
its applications, and relations to other theoretical developments, has been presented \cite{VT13}.

By now we have learned enough about V-T theory to undertake an exact
 formulation of time correlation functions. That is the purpose of the
 present paper. Through linear response theory, time correlation functions
 contain information on nonequilibrium processes (see \cite{HMCDbook}, Ch.~7 and 8).
 Hence in contrast to thermodynamic properties, transits are expected to make
 a significant contribution to time correlation functions. Though we have
 previously studied the velocity autocorrelation function \cite{VT11},
 here the theory will be illustrated by the density autocorrelation
 function $F(q,t)$, since this function probes the atomic motion on a more
 detailed level.
 
 In Sec.~II, formally exact statistical mechanical expressions for the
partition function, equilibrium statistical averages, and time
correlation functions are derived. In Sec.~III, we consider the purely
vibrational contribution to equilibrium statistical averages (when
transits are neglected) and extend the formalism to include time
correlation functions.  The vibrational contribution to $F(q,t)$, and
to its Fourier transform $S(q,\omega)$, are derived in Sec.~IV. 
$S(q,\omega)$ is of interest because it is directly observed
in inelastic neutron and x-ray scattering experiments.  We then
consider a model for incorporating the transit contributions to
$F(q,t)$ and $S(q,\omega)$, and we compare the predictions of V-T
theory with MD calculations.
Our treatment is based on classical statistical mechanics, since this is 
highly accurate for nearly all the elemental liquids, and is appropriate for
comparison with MD. In Sec.~V our conclusions, methods, and aims are related
to the potential energy landscape and mode coupling theory programs,
and we note both significant agreements and divergences.
 We summarize our results and
emphasize the most important conclusions in Sec.~VI.

\section{Statistical averages and time correlation functions}

For simplicity we think of an $N$-atom system in a cubic box, with the
motion governed by periodic boundary conditions. Atom $K$ at time $t$ has
position $\bm{r}_{K}(t)$ and momentum $\bm{p}_{K}(t)$, for $K=1,\dots,N$. The total potential
energy is $\Phi(\{\bm{r}_{K}\})$, and the Hamiltonian is
\begin{equation}\label{eq1}
{\mathcal H}=\sum_{K} \frac{\bm{p}_{K}^{2}}{2M} + \Phi(\{\bm{r}_{K}\}).
\end{equation}
The set $\{\bm{r}_{K}\}$ represents a point in configuration space. If the potential 
valleys are denoted by the index $\gamma$, our decomposition of the potential 
surface means that every accessible configuration is in one and only one
valley $\gamma$. For equilibrium statistical mechanical averages, only the random valleys need
to be considered.

The canonical partition function $\mathcal Z$ is
\begin{equation}\label{eq2}
{\mathcal Z}=\sum_{\gamma}{\mathcal Z}_{\gamma},
\end{equation}
where $\gamma$ is a random valley, and 
\begin{equation}\label{eq3}
{\mathcal Z}_{\gamma} = \frac{1}{h^{3N}}\int d\bm{p}_{1}\dots d\bm{p}_{N}
 \int_{\gamma} d\bm{r}_{1}\dots d\bm{r}_{N}\; e^{-\beta{\mathcal H}}.
\end{equation}
The integral $\int_{\gamma}$ is over the domain of valley $\gamma$, where particle permutations
are not allowed.  The configuration integral is the same in the thermodynamic limit
for each random valley, hence we can define the random valley partition
function ${\mathcal Z}_{rv}$ as ${\mathcal Z}_{\gamma}$ for any random valley $\gamma$. We write the number of
 random valleys as $w^{N}$, where $w$ is a parameter to be determined. Then 
 \begin{equation}\label{eq4}
 {\mathcal Z}=w^{N}{\mathcal Z}_{rv}.
 \end{equation}
 This is exact in the thermodynamic limit. The factor $w^{N}$ gives a 
 contribution $Nk\ln w$ to the entropy, and appears in no other thermodynamic
 function. Calibration to experimental  entropy of melting for the elements
 gives $\ln w = 0.80$ \cite{VT5}.

 Thermodynamic properties such as energy and pressure are given by
equilibrium statistical mechanical averages of the form $\langle A
\rangle$, where $A(\{\mathbf{r}_K, \mathbf{p}_K\})$ is a
time-independent quantity depending on atomic positions and momenta.
With our decomposition of configuration space, the average is
\begin{equation}\label{eq5a}
\langle A \rangle=\frac{\sum_{\gamma}\int d\bm{p}_{1}\dots d\bm{p}_{N}
 \int_{\gamma} d\bm{r}_{1}\dots d\bm{r}_{N}\;A\; e^{-\beta{\mathcal H}}}
          {\sum_{\gamma}\int d\bm{p}_{1}\dots d\bm{p}_{N}
 \int_{\gamma} d\bm{r}_{1}\dots d\bm{r}_{N}\; e^{-\beta{\mathcal H}}}.
\end{equation}
The denominator is proportional to $\mathcal{Z}$, and just as in
Eqs.\ (2-4), every term in the $\sum_\gamma$ in the denominator is
equal, and the same is true of the numerator, with the result
\begin{equation}\label{eq5b}
\langle A \rangle = \langle A \rangle_{rv},
\end{equation}
where $\langle \ldots \rangle_{rv}$ is the average over the domain of a
single random valley.

We now turn to time correlation functions.  Let $B(t)$ be a quantity
depending on atomic positions and momenta as they change in time, so
that $B(t)=B(\{\mathbf{r}_K(t),\mathbf{p}_K(t)\})$.  For complex $B$,
the corresponding autocorrelation function is
\begin{equation}\label{eq5}
C(t) = \langle B(t) B^*(0) \rangle,
\end{equation}
which again invoking the configuration space decomposition becomes
\begin{equation}\label{eq6}
C(t)=\frac{\sum_{\gamma}\int d\bm{p}_{1}\dots d\bm{p}_{N}
 \int_{\gamma} d\bm{r}_{1}\dots d\bm{r}_{N}\;B(t)B^{*}(0)\; e^{-\beta{\mathcal H}}}
          {\sum_{\gamma}\int d\bm{p}_{1}\dots d\bm{p}_{N}
 \int_{\gamma} d\bm{r}_{1}\dots d\bm{r}_{N}\; e^{-\beta{\mathcal H}}},
\end{equation}
where the zero-time variables in $B^*(0)$ are identical with the
integration variables.  When $t=0$, this quantity is a fluctuation, and 
is just a special case of the 
average considered in the previous paragraph, and so
\begin{equation}\label{eq6a}
C(0) = \langle B(0) B^*(0) \rangle_{rv}.
\end{equation}
When $t \neq 0$, however, this is not true, because the system may be
in a different random valley at time $t$ than at time 0, so the
variable $B(t)B^*(0)$ in Eq.~(\ref{eq6}) has to be evaluated along the
equilibrium trajectory of the system.  For this reason the calculation
of time correlation functions involves difficulties that are absent
when computing thermodynamic quantities.

\section{Vibrational Contribution to time correlation functions}

In evaluating the partition function and corresponding equilibrium
thermodynamic functions, we previously developed a tractable model for
the vibrational contribution \cite{VT5}. In this model, the intersections among
valleys are ignored, each random valley is extended to infinite energy, and
averages of the vibrational motion are evaluated for one extended random valley.
The model is applied to time correlation functions in the present Section.

For configurations in random valley $\gamma$, it is useful to write the position of atom $K$ as
\begin{equation} \label{eq8}
\bm{r}_{K}(t) = \bm{R}_{K}(\gamma)+\bm{u}_{K}(t),
\end{equation}
where $\bm{R}_{K}(\gamma)$ is the equilibrium position and $\bm{u}_{K}(t)$ is the displacement from
equilibrium. The set $\{\bm{R}_{K}(\gamma)\}$ is the \textit{structure} $\gamma$. The potential energy in
valley $\gamma$ is denoted $\Phi(\gamma)$, and is expanded about equilibrium as
\begin{equation}\label{eq9}
\Phi(\gamma) = \Phi_{0}(\gamma)+\Phi_{2}(\gamma)+\Phi_{anh}(\gamma),
\end{equation}
where $\Phi_{0}(\gamma)$ is the structure potential, $\Phi_{2}(\gamma)$ is the contribution quadratic
in displacements, and $\Phi_{anh}(\gamma)$ is the remainder of $\Phi(\gamma)$, the anharmonic part. 
The anharmonic contribution is almost always small, often negligible.
The complicated part of $\Phi(\gamma)$ is the presence on its boundary of
intersections with neighboring valleys. But $\Phi(\gamma)$ is a function in $3N$
dimensions, and in most directions there are no such intersections at energies
accessible to the liquid. It is  only a few directions where 
low lying valley-valley intersections are present, and it is only in these
few directions where transits occur. Hence to describe the vibrational
motion alone, we shall ignore intervalley intersections, and as a leading
 approximation we shall also neglect anharmonicity. The corresponding 
 \textit{extended} random valley has potential energy
\begin{equation}\label{eq10}
\Phi = \Phi_{0} + \frac{1}{2} \sum_{KLij} \Phi_{Ki,Lj} u_{Ki} u_{Lj}, \;\;\;\;\;\;\; 0\leq u_{Ki}<\infty.
\end{equation}
The notation is $i$, $j =$ Cartesian component $x,y,z$, and $ \Phi_{Ki,Lj}$ are
second position derivatives of $\Phi$ at equilibrium. The index $\gamma$ is suppressed in Eq.~(\ref{eq10}).

The Hamiltonian  for motion in an extended random valley is denoted
${\mathcal H}_{vib}$, for vibrations, and is
\begin{equation}\label{eq11}
{\mathcal H}_{vib} = \Phi_{0} + \sum_{Ki} \frac{p_{Ki}^{2}}{2M} + 
\frac{1}{2} \sum_{KLij} \Phi_{Ki,Lj} u_{Ki} u_{Lj}. 
\end{equation}
It is useful to transform from displacements to normal modes of vibration.
The normal modes are labeled $\lambda=1,\dots,3N$, their coordinates are $Q_{\lambda}$, and
the transformation of displacements is
\begin{equation}\label{eq12}
u_{Ki}(t)= \sum_{\lambda} w_{Ki,\lambda}\; Q_{\lambda}(t),
\end{equation}
where for each $\lambda$, $w_{Ki,\lambda}$ are $3N$ real components of eigenvector $\lambda$. The
eigenvectors diagonalize the matrix of potential energy coefficients $\Phi_{Ki,Lj}$,
and they satisfy
\begin{equation}\label{eq13}
\sum_{Ki} w_{Ki,\lambda} w_{Ki,\lambda'} = \delta_{\lambda \lambda'},\;\;\; \mbox{orthonormality;}
\end{equation}
\begin{equation}\label{eq14}
\sum_{\lambda} w_{Ki,\lambda} w_{Lj,\lambda} = \delta_{KL}\delta_{ij}, \;\;\;\mbox{completeness.}
\end{equation}
For a given extended random valley, its microscopic geometry is encoded in
its eigenvectors and its structure. The Hamiltonian (\ref{eq11}) transforms to 
\begin{equation}\label{eq15}
{\mathcal H}_{vib}=\Phi_{0}+ \sum_{\lambda} \left[\frac{P_{\lambda}^{2}}{2M} + 
\frac{1}{2} M\omega_{\lambda}^{2}Q_{\lambda}^{2}\right],
\end{equation}
where $P_{\lambda}=M\dot{Q}_{\lambda}$, and where $\omega_{\lambda}$ is the vibrational frequency of mode $\lambda$.
By definition, $\omega_{\lambda}^{2}>0$ for $3N-3$ modes, and $\omega_{\lambda}^{2}=0$ for the three modes
of uniform translation. It is understood that the translational modes
are omitted from statistical mechanical analyses.  The structural potential
$\Phi_{0}$, and the distribution $g(\omega)$ of normal mode frequencies, are each the
same for all random valleys of a given material.

The partition function is written in Eq.~(\ref{eq4}). The vibrational contribution is
 ${\mathcal Z}_{vib}=w^{N}{\mathcal Z}_{hrv}$, where ${\mathcal Z}_{hrv}$ is the partition function for an extended harmonic
  random valley. The result is
\begin{equation}\label{eq16}
{\mathcal Z}_{vib}=w^{N} e^{-\beta\Phi_{0}}\prod_{\lambda}\frac{kT}{\hbar\omega_{\lambda}}.
\end{equation}
This is subject to a correction for anharmonicity, and a correction to recover the
proper domain of integration for ${\mathcal Z}_\gamma$, or ${\mathcal Z}_{rv}$. The latter is called the boundary
correction. Comparison with experimental data shows that both corrections are small for elemental liquids at melt. 

The time correlation function $C(t)$ is written in Eq.~(\ref{eq6}). The vibrational
contribution is $C_{vib}(t)$, obtained by replacing each random valley by its extension. The system then remains
in a single random valley, and Eq.~(\ref{eq6}) reduces to
\begin{equation}\label{eq17}
C_{vib}(t)=\left < B(t) B^{*}(0) \right >_{hrv},
\end{equation}
where $\left<\dots\right>_{hrv}$ is the average for an extended harmonic random valley. The average
in Eq.~(\ref{eq17}) can be resolved into normal-mode time correlation
functions.  In quadratic order these functions are
\begin{equation}\label{eq18}
\left< Q_{\lambda}(t) Q_{\lambda'}(0) \right>_{hrv}= \frac{kT}{M\omega_{\lambda}^{2}} \delta_{\lambda\lambda'}\cos\omega_{\lambda}t,\\
\end{equation}
\begin{equation}
\label{eq19}
\left< P_{\lambda}(t) \;P_{\lambda'}(0) \right>_{hrv}= MkT \delta_{\lambda\lambda'}\cos\omega_{\lambda}t,\\
\end{equation}
\begin{equation}
\label{eq20} 
\left< Q_{\lambda}(t)\; P_{\lambda'}(0) \right>_{hrv}= 0.
\end{equation}
An exemplary time correlation function is studied in the next Section.

\section{The Density Autocorrelation Function}

The density autocorrelation function, or intermediate scattering function, 
is $F(q,t)$,
\begin{equation}\label{eq21}
F(q,t) = \frac {1}{N} \left < \rho (\bm{q},t) \rho (-\bm{q},0) \right >,
\end{equation}
where $\rho (\bm{q},t)$ is the Fourier transform of the density operator,
\begin{equation}\label{eq22}
\rho (\bm{q},t) = \sum_{K} e^{-i \bm{q}\cdot \bm{r}_{K}(t)}.
\end{equation}
First we want to evaluate the vibrational contribution to $F(q,t)$. From Eq.~(\ref{eq17}), this is 
\begin{equation}\label{eq23}
F_{vib}(q,t) = \frac {1}{N} \left < \rho (\bm{q},t) \rho (-\bm{q},0) \right >_{hrv}.
\end{equation}
Because an equilibrium liquid is macroscopically isotropic, the average in (\ref{eq21}),
or in (\ref{eq23}), cannot depend on the angle of $\bm{q}$ in the thermodynamic limit. But we
need to treat a finite system, for which numerical evaluations are possible. 
For a finite cubic system with periodic boundary conditions, the $\bm{q}$ form a
discrete allowed set of wavevectors consistent with the periodicity. We are therefore 
allowed to average over the directions of the allowed $\bm{q}$.
Then Eq.~(\ref{eq23}) becomes
\begin{equation}\label{eq24}
F_{vib}(q,t) = \frac {1}{N} \left<\left < \rho (\bm{q},t) \rho (-\bm{q},0) \right >_{hrv}\right>_{\bm{q}^{\ast}},
\end{equation}
where $\left< \dots \right>_{\bm{q}^{\ast}}$ is the average over the star of $\bm{q}$.

For motion in an extended random valley, $\bm{r}_{K}(t)$ is written from Eq.~(\ref{eq8}), and
the average in   Eq.~(\ref{eq24}) works out as (the algebra is the same as for the crystal \cite{Lovbook,Glybook})
\begin{eqnarray} \label{eq25}
\lefteqn{\left < \rho (\bm{q},t) \rho (-\bm{q},0) \right >_{hrv}}\nonumber \\
& & = \sum_{KL} e^{-i \bm{q}\cdot \bm{R}_{KL}} 
 \left < e^{-i \bm{q} \cdot (\bm{u}_{K}(t) - \bm{u}_{L}(0))} \right >_{hrv} \nonumber \\
& &  =	 \sum_{KL} e^{-i \bm{q}\cdot \bm{R}_{KL}} 	
   e^{-\frac{1}{2} \left <
           \left [ \bm{q}\cdot(\bm{u}_{K}(t)-\bm{u}_{L}(0)) \right ]^{2} \right >_{hrv}},
\end{eqnarray}
where $\bm{R}_{KL}=\bm{R}_{K}-\bm{R}_{L}$. The displacement averages
simplify to
\begin{eqnarray}\label{eq26}
\lefteqn{-\frac{1}{2} \left<\left[
                \bm{q} \cdot (\bm{u}_{K}(t) - \bm{u}_{L}(0)) \right ]^{2}\right >_{hrv} =} \nonumber\\
& &-W_{K}(\bm{q}) - W_{L}(\bm{q}) 			
				+\left < \bm{q}\cdot\bm{u}_{K}(t) \;\bm{q} \cdot\bm{u}_{L}(0) \right  >_{hrv}.
\end{eqnarray}
$W_{K}(\bm{q})$ is the Debye-Waller factor for atom K,
\begin{equation}\label{eq27}
W_{K}(\bm{q})=\frac{1}{2}\left < (\bm{q}\cdot\bm{u}_{K})^{2}\right  >_{hrv}.
\end{equation}
The time dependence is entirely contained in the displacement-displacement
correlation functions $\left < \bm{q}\cdot\bm{u}_{K}(t) \;\bm{q} \cdot\bm{u}_{L}(0) \right  >_{hrv}$. For a random valley, these
functions vanish as $t\rightarrow\infty$. Hence it is convenient to write $F(q,t)$ in the form
\begin{equation}\label{eq28}
F_{vib}(q,t)=F_{vib}(q,\infty)+\left[F_{vib}(q,t)-F_{vib}(q,\infty) \right],
\end{equation}
where
\begin{equation} \label{eq29}
F_{vib}(q,\infty) = \frac {1}{N} \left < \sum_{KL}\cos \bm{q}\cdot \bm{R}_{KL} e^{-W_{K}(\bm{q})} e^{-W_{L}(\bm{q})}
                     \right >_{\bm{q}^{\ast}},
\end{equation}
\begin{eqnarray}\label{eq30}
\lefteqn{F_{vib}(q,t)-F_{vib}(q,\infty)=
\frac {1}{N} \Bigg < \sum_{KL}\cos \bm{q}\cdot \bm{R}_{KL}}\nonumber \\
& &\times\; e^{-W_{K}(\bm{q})}e^{-W_{L}(\bm{q})}
\;\left[ e^{\left < \bm{q} \cdot \bm{u}_{K}(t)\bm{q} \cdot \bm{u}_{L}(0)\right >_{hrv}}-1\right]
                     \Bigg >_{\bm{q}^{\ast}} .
\end{eqnarray}
Notice that $F_{vib}(q,\infty)$ is positive definite.

The dynamic structure factor $S(q,\omega)$ is directly observed in neutron and
 photon scattering experiments, and is the Fourier transform of $F(q,t)$,
 \begin{equation}\label{eq31}
S(q,\omega)=\frac{1}{2\pi}\int_{-\infty}^{\infty} F(q,t) e^{i\omega t}dt.
\end{equation}
The transform will be evaluated in the one-mode approximation, obtained by 
expanding $e^{\left<\dots\right>\;_{hrv}}-1$ to leading order in Eq.~(\ref{eq30}), and denoted by $F_{vib}^{(1)}(q,t)$
in place of  $F_{vib}(q,t)$:
\begin{eqnarray}\label{eq32}
\lefteqn{F_{vib}^{(1)}(q,t)-F_{vib}(q,\infty)=
\frac {1}{N} \Bigg < \sum_{KL}\cos \bm{q}\cdot \bm{R}_{KL}  }\nonumber\\
& &\times\;e^{-W_{K}(\bm{q})} e^{-W_{L}(\bm{q})}
\;\left < \bm{q} \cdot \bm{u}_{K}(t)\;\bm{q} \cdot \bm{u}_{L}(0)\right >_{hrv}
                     \Bigg >_{\bm{q}^{\ast}} .
\end{eqnarray}
The time correlation function is transformed by Eqs.~(\ref{eq12}) and (\ref{eq18}) to give
 \begin{eqnarray}\label{eq33}
\lefteqn{\left<\bm{q}\cdot\bm{u}_{K}(t)\;\bm{q}\cdot\bm{u}_{L}(0)\right>_{hrv} =} \nonumber \\
& &\sum_{\lambda} \frac{kT}{M\omega_{\lambda}^{2}}\bm{q}\cdot \bm{w}_{K\lambda}\; \bm{q}\cdot \bm{w}_{L\lambda}
\cos\omega_{\lambda}t.
\end{eqnarray}
Then $S(q,\omega)$ in the one-mode approximation is 
\begin{equation}\label{eq34}
S_{vib}(q,\omega) = F_{vib}(q,\infty)\delta(\omega)+S_{vib}^{(1)}(q,\omega),
\end{equation}
where 
\begin{equation} \label{eq35}
S_{vib}^{(1)}(q,\omega)=\frac{3kT}{2M}\frac{1}{3N}\sum_{\lambda}f_{\lambda}(q)[\delta(\omega+\omega_{\lambda})+
\delta(\omega-\omega_{\lambda})], 
\end{equation}
\begin{eqnarray} \label{eq36}
\lefteqn{f_{\lambda}(q)=\frac{1}{\omega_{\lambda}^{2}}\Bigg < \sum_{KL}\cos \bm{q}\cdot \bm{R}_{KL} } \nonumber\\
& &\times\;e^{-W_{K}(\bm{q})} e^{-W_{L}(\bm{q})}
\;\ \bm{q} \cdot \bm{w}_{K\lambda}\;\bm{q} \cdot \bm{w}_{L\lambda}
                     \Bigg >_{\bm{q}^{\ast}}.
\end{eqnarray}
Finally the normal mode resolution of the Debye-Waller factor is
\begin{equation}\label{eq37}
W_{K}(\bm{q})=\sum_{\lambda}\frac{kT(\bm{q}\cdot\bm{w}_{K\lambda})^{2}}{2M\omega_{\lambda}^{2}}.
\end{equation}

Some physical interpretation of $S_{vib}(q,\omega)$ is helpful. In Eq.~(\ref{eq34}), 
$F_{vib}(q,\infty)\delta(\omega)$ is the elastic scattering cross section at momentum transfer ${\hbar}q$. 
Eq.~(\ref{eq35}) expresses $S_{vib}^{(1)}(q,\omega)$ as the sum over all modes of the one-mode cross section for scattering
at momentum transfer ${\hbar}q$, where  $f_{\lambda}(q)\delta(\omega+\omega_{\lambda})$ is proportional to the cross 
section for scattering with annihilation of energy $\hbar\omega_{\lambda}$ in mode $\lambda$, and 
$f_{\lambda}(q)\delta(\omega-\omega_{\lambda})$ is proportional to the cross 
section for scattering with creation of energy $\hbar\omega_{\lambda}$ in mode $\lambda$. 
The proportionality constant consists of the factors in front of $\sum_{\lambda}$ in Eq.~(\ref{eq35}).
One observes from Eq.~(\ref{eq36}) that $f_{\lambda}(q)>0$. For
most elemental liquids at $T{\approx}T_{m}$, we expect the one-mode approximation to provide 
a reasonable first approximation over the range of $q$ 
where vibrational scattering is important.
What is missing from the one-mode approximation is vibrational anharmonicity and multi-mode scattering events.

We now allow for transits (the following argument is also found in \cite{ARXIV05m}).  
 When atom $K$ is involved in a transit,
both $\bm{R}_{K}$ and $\bm{u}_{K}$ change in a very short time, in such a way that 
$\bm{R}_{K} + \bm{u}_{K}$ 
remains continuous and differentiable in time. A detailed model of transits
in the atomic trajectory may be found in Chisolm et al. \cite{VT11,VT13}. Here we
seek a simpler approximation. If the time segments between transits involving
atom $K$ are denoted $\gamma_{K}=1,2,\dots$, then the position of atom $K$ at time $t$ is
$\bm{R}_{K}(\gamma_{K}(t)) + \bm{u}_{K}(\gamma_{K}(t),t)$. $F(q,t)$ for the liquid is then
written, from Eq.~(\ref{eq21}),
\begin{equation} \label{eq40}
F_{liq}(q,t) = \frac{1}{N}
 \left <\sum_{KL} e^{-i \bm{q}\cdot [\bm{R}_{K}(\gamma_{K}(t))-\bm{R}_{L}(t=0)]} 
\;e^{-i \bm{q} \cdot [\bm{u}_{K}(\gamma_{K}(t),t) - \bm{u}_{L}(t=0)]} \right >.
\end{equation}
Our numerical studies provide evidence, described below in connection with
Eq.~(\ref{eq43}), that transits can be approximately neglected in the displacements
$\bm{u}_{K}(\gamma_{K}(t),t)$. We therefore make this approximation, and separately
average the displacement terms in Eq.~(\ref{eq40}) over harmonic vibrations. The 
harmonic average can be simplified to give the result,
\begin{equation}  \label{eq41}
 F_{liq}(q,t) \approx
\frac {1}{N} \Bigg < \sum_{KL}e^{-i \bm{q}\cdot [\bm{R}_{K}(\gamma_{K}(t))-\bm{R}_{L}(0)]}
\; e^{-W_{K}(\bm{q})}e^{-W_{L}(\bm{q})}
\;\left[ 1 + \left < \bm{q} \cdot \bm{u}_{K}(t)\;\bm{q} \cdot \bm{u}_{L}(0)\right >_{h}+ \dots\right]
                     \Bigg >_{\bm{q}^{\ast}} .
\end{equation}
Our next step is to modify this equation so as to model the presence of transits in 
$\bm{R}_{K}$.

There are two ways in which transits contribute to $F_{liq}(q,t)$. First,
transits introduce a fluctuating phase in the complex exponential in Eq.~(\ref{eq41}),
and this causes additional time decay through decorrelation along each
atomic trajectory. We model this with a relaxation function of the form
$e^{-\alpha t}$. Second, transits give rise to inelastic scattering, in addition to the
vibrational mode scattering already present in Eq.~(\ref{eq41}), and this increases
the total scattering cross section. We model this with a multiplicative
factor.

The leading term in Eq.~(\ref{eq41}) gives rise to the liquid Rayleigh peak, and
so is denoted $F_{R}(q,t)$. Without transits, $F_{R}(q,t)$ reduces to $F_{vib}(q,\infty)$,
Eq.~(\ref{eq29}), so we model $F_{R}(q,t)$ as 
\begin{equation} \label{eq42}
F_{R}(q,t) = C(q)\; F_{vib}(q,\infty)\; e^{-\alpha_{1}(q)\; t}.
\end{equation}
This function decays to zero with increasing time, in accord with the liquid
property $F_{liq}(q,t) \rightarrow 0$ as $t \rightarrow \infty$. 
The relaxation rate $\alpha_{1}(q)$ is expected to
be around the mean single-atom transit rate. $C(q)$ is positive, and greater than $1$
because of the inelastic scattering associated with transits (notice the total scattering
cross section is not affected by the factor $e^{-\alpha_{1} t}$).

The displacement-displacement correlation function in Eq.~(\ref{eq41}) gives rise to
the Brillouin peak, and so is denoted $F_{B}(q,t)$. Without transits $F_{B}(q,t)$
reduces to $F_{vib}(q,t)-F_{vib}(q,\infty)$, Eq.~(\ref{eq32}). Empirically, we have found that
this vibrational contribution alone gives an excellent account of the location
of the Brillouin peak, and the total cross section within it, as compared
with MD calculations and with experimental data for liquid sodium \cite{ARXIV05}. 
This suggests  keeping the vibrational contribution intact, as we did in
going to Eq.~(\ref{eq41}), and also suggests that we model $F_{B}(q,t)$ by

\begin{equation} \label{eq43}
F_{B}(q,t) = [F_{vib}(q,t) - F_{vib}(q,\infty)]\;e^{-\alpha_{2}(q) \;t}.
\end{equation}
Note $F_{vib}(q,t) - F_{vib}(q,\infty)$ decays to zero with time, because of the decay of
the vibrational correlation function in Eq.~(\ref{eq32}), and this decay gives the  
Brillouin peak its natural width \cite{ARXIV05}. The right side of Eq.~(\ref{eq43})
decays faster with time, hence broadens the Brillouin peak from its natural width,
but leaves its total cross section unchanged.

From the above equations, our model for the dynamic structure factor is
\begin{equation} \label{eq44}
S_{liq}(q,\omega)=S_{R}(q,\omega)+S_{B}(q,\omega),
\end{equation}
\begin{equation}\label{eq45}
S_{R}(q,\omega) = \frac{C(q)\; \alpha_{1}\; F_{vib}(q,\infty)}{\pi(\omega^{2}+\alpha_{1}^{2})},
\end{equation}
\begin{equation}\label{eq46}
S_{B}(q,\omega) = \frac{3kT}{2M}\frac{1}{3N}\sum_{\lambda}f_{\lambda}(q)\frac{1}{\pi}
\left [ \frac{\alpha_{2}}{(\omega +\omega_{\lambda})^{2}+\alpha_{2}^{2}} +
 \frac{\alpha_{2}}{(\omega -\omega_{\lambda})^{2}+\alpha_{2}^{2}}\right ].
\end{equation}
The model has three adjustable parameters for each $q$, namely $C(q), \alpha_{1}(q)$, and
$\alpha_{2}(q)$. For a discussion of how we determine the values of these parameters, as well as other details
of the calculation such as the addressing of possible short-time errors, see \cite{ARXIV05m}.

\begin{figure}[h]
\includegraphics[height=3.0in,width=3.0in,keepaspectratio]{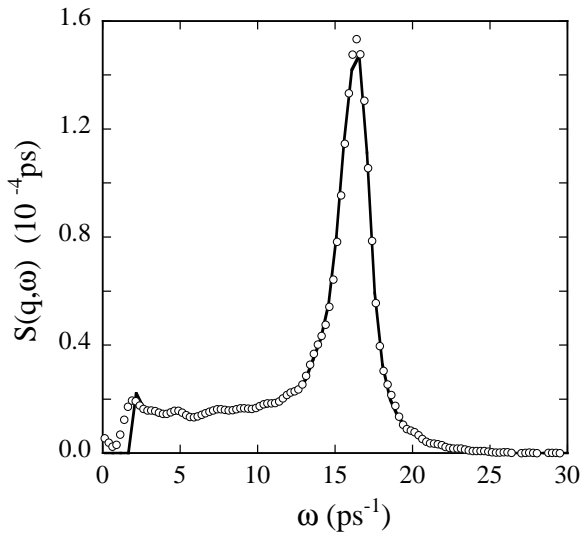}
\caption {\label{Fig2}  $S(q,\omega)$ for motion in a single random valley at $q=0.29711\,a_{0}^{-1}$
at 24~K: solid line is  harmonic vibrational theory (Eqs.~(\ref{eq35}-\ref{eq37})), circles are MD.}
\end{figure}

\begin{figure}[h]
\includegraphics[height=3.0in,width=3.0in,keepaspectratio]{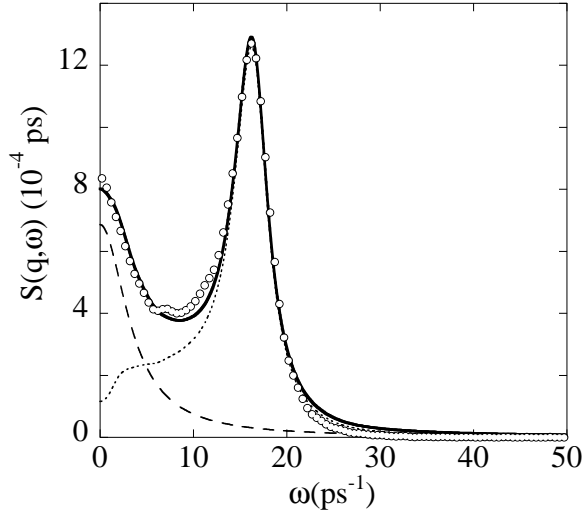}
\caption {\label{S112} $S(q,\omega)$ for $q=0.29711$~a$_{0}^{-1}$ at 395~K, from the model (solid line)
 and from MD (circles). The Rayleigh (broken line) and the Brillouin (dotted line) contributions 
 to the model are also shown separately.}
\end{figure}

At this point let us compare theory with MD. We have done calculations for a system with
$N=500$ atoms and an interatomic potential representing metallic sodium at the density of 
the liquid at melt. The potential gives an accurate account of the vibrational and thermodynamic
properties of crystal and liquid phases, and a good account of self diffusion in the liquid 
(for summaries see \cite{VT15,VT13}). When the MD system is quenched to 24~K, transits do not occur,
and the equilibrium system moves only within a single random valley. The circles in Fig.~\ref{Fig2}
show the corresponding
$S(q,\omega)$ from MD, calculated directly from the defining equations (\ref{eq21}), (\ref{eq22}), and 
(\ref{eq31}) at $q=0.29711\,a_{0}^{-1}$ ($=0.28\,q_{max}$ where $q_{max}=1.05\,a_{0}^{-1}$ is the position
of the first peak of the static structure factor).
The theoretical equations (\ref{eq35}-\ref{eq37}) for $S_{vib}^{(1)}(q,\omega)$ were evaluated for 
the same temperature and the same random valley, where the $\delta$-functions were smoothed 
by making a histogram, and the results are shown by the solid line in Fig.~\ref{Fig2}. As the Figure shows,
the theory reproduces the Brillouin peak from the MD results perfectly.  $S_{vib}^{(1)}(q,\omega)$ is 
zero at low frequencies because our system has no normal mode frequencies below 1.7 ps$^{-1}$.  
From Eq.~(\ref{eq34}), elastic scattering contributes the peak $F(q,\infty)\delta(\omega)$ in theory and MD alike.
This peak is not indicated in Fig.~\ref{Fig2}.  To test the transit model, we then ran the MD system at 395~K,
at which temperature the system has melted and is transiting frequently; the resulting $S(q,\omega)$ 
 at the same $q$ is shown by the circles in Fig.~\ref{S112}.  We also evaluated 
 equations (\ref{eq44}-\ref{eq46}) of the model at the same temperature; for details of the evaluation,
including the suppression of finite-$N$ effects and the criteria used to determine the three parameters, 
see \cite{ARXIV05m}.  The individual functions $S_{R}(q,\omega)$, $S_{B}(q,\omega)$, and their sum $S_{liq}(q,\omega)$ 
are shown as the broken, dotted, and solid lines in the Figure, respectively.  Just as the purely vibrational
theory accounted superbly for the Brillouin peak at low temperatures, the inclusion of the model for transits
describes the Rayleigh and Brillouin peaks very accurately.  Notice that the location of the Brillouin peak has not
shifted from the location predicted by the vibrational part, as predicted earlier.  For additional calculations at
different $q$ and a detailed discussion of such matters as the importance of multimode scattering and the 
significance of the magnitudes of the model parameters, see \cite{ARXIV05m}.

\section{Comparison of theories}

An extensive body of work on the dynamics of liquids has been performed in the potential energy landscape
and mode coupling theory research programs. In this Section we discuss those programs and compare
our aims and results with theirs.

\subsection{Potential Energy Landscape  (PEL) Theories}

Instantaneous normal modes (INM) were first reported in the MD calculations of Rahman et al.~\cite{E7}.
These are the vibrational modes which diagonalize the potential
energy curvature tensor at an arbitrary point on the potential energy surface. INM data
are obtained by averaging over randomly chosen configurations on an equilibrium MD
trajectory. Seeley and Keyes \cite{AA} argued that the INM averaged density of states
contains information which may be used to construct theories of liquid state
dynamics. The idea was applied to self diffusion \cite{AB,AC,AE,AG,AH} by using
INM data to estimate the hopping rate for Zwanzig's approximation \cite{E8}.
Bembenek and Laird \cite{CD,CE} classified the imaginary frequency INM as unstable 
modes (double-well modes) and shoulder modes, and corresponding improvements
in the fitting of calculated velocity autocorrelation functions followed \cite{AI,AJ,AM,AN}.
Buchner, Ladanyi, and Stratt \cite{BA} started with the Hamiltonian for harmonic
vibrations in a system of diatomic molecules, and showed that the INM
description is accurate only for very short times, but that reasonable agreement
is mantained for longer times if the imaginary frequency modes are omitted.
A mean field theory was studied \cite{CA,CB,BB}, as well as time evolution and
mixing of INM \cite{BF}. Kr{\"a}mer et al.~\cite {CG} showed that INM data at closely
spaced points on the MD trajectory will give accurate results for the MD time
correlation functions. To investigate the plane wave character of sound modes in liquid ZnCl$_{2}$,
Ribeiro, Wilson and Madden \cite{CF} carried out an INM
calculation of $S(q,\omega)$. This application of INM theory treats  instantaneous
configurations as if they were equilibrium configurations, hence the vibrational
motion used is not a solution of the equations of motion (see Keyes and Fourkas \cite{AO}).
Additional INM calculations have been done for Rb \cite{CC}, for Na \cite{CH,CI,CJ,CK},
and for water \cite{BD,HF,HG,HH}.

In contrast to INM, quenched normal modes (QNM) are the vibrational modes
evaluated at a local potential energy minimum (an equilibrium configuration). QNM
data are usually obtained by averaging over the potential energy minima sampled
from an equilibrium MD trajectory. For water at 300~K, Pohorille et al.~\cite{DA} found 
that the radial distribution calculated from QNM data are very similar to the
MD results. Application of QNM theory to water was reviewed by Ohmine and Tanaka \cite{DB}.
Cao and Voth \cite{DC,DD} argued that, because of the thermal
 fluctuations, each potential energy valley must be replaced by a self-consistent
 temperature-dependent mean field. Notice this precludes a Hamiltonian
 formulation of liquid dynamics, because a Hamiltonian requires a time
 and temperature independent many-body potential. Several authors \cite{DEa,DEb,DEc} discussed the issue  of
 whether or not local information on the potential surface, embodied by the 
 distribution of unstable instantaneous normal modes, can be used to predict the
 hopping rates and barrier heights for Zwanzig's model \cite{E8} of self diffusion. Rabani, Gezelter and
 Berne \cite{DF} introduced the cage correlation function, which
 measures the rate of change of atomic surroundings, and combined this with
 Zwanzig's approximation to account for diffusion in LJ systems. For CS$_{2}$,
 these authors \cite{DG} concluded that neither the INM nor the QNM density of states
 is the correct one to use in Zwanzig's approximation. In the language of V-T
 theory, the cage correlation function aims to account for the mean transit-induced 
 hopping motion of a single atom.
 
 Currently the inherent structure (IS) research program is directed mainly toward
 supercooled liquids and the glass transition. The standard technique is to probe
 the potential energy surface by quenching at random times along the equilibrium
 MD trajectory for glass forming systems.  The data collected this way are
 equilibrium statistical averages and depend on the temperature of the MD
 calculation. Systems studied include binary LJ mixtures, models for water,
 SiO$_{2}$, and orthoterphenyl. The binary LJ systems apparently do not crystallize,
 but do exhibit liquid behavior above an ``effective" melting temperature $T_{m}$.
 For the Kob-Andersen binary system \cite{LA}, for example, $T_{m}\sim 1.0$, well above
 the mode-coupling critical temperature $T_{MCT}=0.435$ \cite{FC}. The following
 picture emerges. The average inherent structure energy is virtually constant
 for $T\gtrsim T_{m}$, but decreases as $T$ decreases below $T_{m}$ \cite{FA,FB,FC,FE,FF,HG,IA,JC}.
 Schr{\o}der et al. \cite{JH} presented numerical evidence that, below a
 crossover temperature in the vicinity of $T_{MCT}$, the system dynamics can be
 separated into vibrations around inherent structures and transitions between
 them. Potential energy saddle analyses also reveal this separation of system
 dynamics \cite{IA,IC,ID}. Consensus is that the MD systems are in thermodynamic
 equilibrium at temperatures down to $T_{MCT}$. Structural entropy decreases
 toward zero as temperature decreases toward $T_{MCT}$ \cite{FA,FC,FF,HA,HB,HC,HG}.
 The precise character of this decrease in structural entropy depends on the
 temperature-dependent \textit{distribution} of the inherent structure energies \cite{FA,
 FB,FC,FD,FE}. The Gaussian character of this distribution at low temperatures has
 been established for binary LJ systems by numerical analysis, and has been
 shown on rather general arguments to result from the central limit theorem as
 $N\rightarrow \infty$ \cite{S4}. Numerical confirmation of the Adams-Gibbs relation \cite{E9} between
 the decrease in structural entropy and the slowing down of structural relaxation 
 has been obtained for several systems \cite{HA,HB,HC,HD}. The key role
 of Adams-Gibbs in the IS theory of the glass transition is discussed by Shell and Debenedetti \cite{HE}.
 Recent work on ageing of the glass is of interest but beyond our 
 present scope \cite{JD,JG,JI,JK}.

\subsection{Comparison of V-T and PEL theories}

The goal of PEL theories is to describe statistical mechanical properties of
liquids and supercooled liquids in terms of temperature-dependent statistical
mechanical properties of the potential energy surface. INM and QNM theories
work with temperature-dependent averages of vibrational mode data, and seek 
to explain time correlations such as velocity and density autocorrelations. IS
theories work with temperature-dependent averages of structural potential energies
and vibrational frequencies, and seek to explain such quantities as
structural and vibrational entropy. In contrast, the goal of V-T theory is to
construct a tractable and reasonably accurate zeroth order Hamiltonian, from which
both equilibrium thermodynamic properties and time correlation functions can be
calculated without adjustable parameters, and then proceed with Hamiltonian
corrections to the zeroth order theory. Ultimately this is the most basic and
comprehensive formulation available to theoretical physics.

The free energy in IS theory contains the important contribution $f_{basin}$, the
free energy of the liquid constrained to be in one of its characteristic basins. The
difficulties posed by this constraint are examined in detail by Sastry \cite{FF}, and
Sciortino observes that there is no unique way to enforce the constraint \cite{S3}.
In contrast, V-T theory incorporates this constraint from the outset \cite{VT5}, in a 
manner consistent with all available experimental data, by replacing each random
valley by its harmonic extension to infinity, with valley-valley intersections
ignored. This formulation naturally separates in leading order and at all temperatures the intravalley
vibrational motion and the intervalley transit motion \cite{VT5}, a separation later
found in MD calculations at temperatures in the vicinity of $T_{MCT}$ \cite{JH,IA,IC,ID}.
The V-T formulation also separates in leading order and at all temperatures the harmonic vibrational free
energy from other free energy contributions, a separation again found in MD 
calculations for temperatures below melting \cite{FA,FC}.

To test the predicted uniformity \cite{VT5} of random valleys, we calculated
the structural potential per atom and the vibrational frequency averages $\left<\omega\right>,
\left<\omega^{2}\right>$ and $\left<\ln(\omega/\omega_{0})\right>$ for several
monatomic random valleys with $N=500-3000$ \cite{VT8,VT9}.
The rms scatter was well below $0.01kT_{m}$ for the structural potential per atom, and was
about $0.01\%$ of the mean for the vibrational frequency averages \cite{VT8,VT9}. This scatter
is negligible in V-T theory for $T\gtrsim T_{m}$. We also presumed that the scatter would
vanish as $N\rightarrow\infty$. This limiting property is verified for the structural potential
per atom by the analysis of Heuer and B\"{u}chner (see Eq.~(14) with $M\rightarrow N$ of \cite{S4}).

As PEL theories develop, they are increasingly concerned with anharmonicity as
it affects the interrelations among statistical mechanical averages \cite{BE,BF,FG,FH}.
In the PEL approach, anharmonicity is taken to mean all atomic motion which is not 
harmonic vibrational motion. Anharmonicity therefore poses severe theoretical
difficulties as soon as transits become important, i.e. at $T$ a little above $T_{MCT}$.
Sciortino observes that there is currently no satisfactory model for anharmonicity at
$T$ above $T_{MCT}$ \cite{S3}. In contrast, V-T theory separates the correction to the zeroth order 
potential energy surface into two distinct contributions \cite{VT5}: (a) the boundary
correction which subtracts the potential energy surfaces that were extended 
beyond the intersections of valleys, and (b) the anharmonic correction which is the
anharmonic potential of each random valley up to its intersection with a neighboring valley.
This is a most useful separation because the boundary correction to entropy 
is always negative, can be reasonably estimated, and is
of increasing importance as temperature increases \cite{VT5,VT6}. The
anharmonic free energy is quite complicated but, in our experience so far, it is always
small in the liquid \cite{VT5,VT14}, just as it is in the crystal \cite{VT15}.

Recent MD studies of complex systems exhibit behaviors characteristic of the V-T
theory description of the potential surface.  For a model of orthoterphenyl, the 
anharmonic entropy is very small at all $T$, the structural entropy is small and has
very little temperature dependence, and the harmonic vibrational entropy is most
of the total (85\%) and has nearly all the temperature dependence (Fig.~12 of \cite{FG}).
These properties are very similar to the entropy contribution for all monatomic
elemental liquids, as expressed in V-T theory \cite{VT5,VT6}. In addition, the excess of the
mean potential energy over the harmonic vibrational contribution decreases for the
binary LJ system at $T\gtrsim T_{m}$ (Fig.~5 of \cite{IB}). This decrease is
present in all monatomic elemental liquids at $T\gtrsim T_{m}$, where it has been attributed to the 
boundary contribution \cite{VT5, VT6}.

That the random valleys dominate the statistical mechanics of monatomic 
liquids at $T\gtrsim T_{m}$ has been demonstrated by MD calculations (Figs.~1 and 2 of \cite{VT8}).
Within the results of recent MD calculations for systems more complex than
the monatomics, there is also strong evidence for the statistical mechanical
dominance of macroscopically uniform (hence random) valleys at $T\gtrsim T_{m}$. The
most prominent such evidence  is (a) the mean inherent structure energy has very
little temperature dependence at each fixed density for $T\gtrsim T_{m}$, and (b) the mean vibrational frequency
distribution and its log moment are nearly independent of temperature at each fixed density for $T\gtrsim T_{m}$.
Property (a) is found in monatomic LJ (Fig.~2 of \cite{IA}), in binary LJ systems (Fig.~1A of \cite{FA}),
Fig.~2A lower of \cite{FB},
Fig.~4 top of \cite{FC},
Fig.~8 of \cite{FE},
Fig.~2 of \cite{FF},
Fig.~5 of \cite{IB},
Fig.~1a of \cite{JC},
Fig.~1a of \cite{JK})
and in a model for water (Fig.~1b of \cite{HG}). Property (b) is found in
binary LJ systems (Figs.~6 and 7 of \cite{FB}, Fig.~10 of \cite{FE}, Figs.~2a and 3 of \cite{JK})
 and in the model for water (Figs.~1b and 2c of \cite{HG}). It therefore appears
that the symmmetry classification of potential energy valleys as random and
symmetric, with their associated properties, as introduced in V-T theory \cite{VT5}
is appropriate for liquids more complex
than monatomics.

\subsection{Comparison of V-T and  Mode Coupling theories}

Detailed summaries of mode coupling theories of liquid
dynamics are given by Boon and Yip \cite{BYbook}, Hansen and McDonald \cite{HMCDbook}, and
Balucani and Zoppi \cite{BZbook}.  Applied to the glass transition, mode coupling theory predicts a dynamic
singularity when temperature is lowered below the critical temperature $T_{MCT}$ \cite{MB,
MC}. From MD calculations for the binary LJ system, Kob and Andersen \cite{LA,LB,
LC} showed that mode coupling theory is able to rationalize the density correlation
functions at low temperatures, in the vicinity of $T_{MCT}$. Further tests of mode coupling theory for
glassy dynamics are reviewed by G\"{o}tze \cite{ME}. Mode coupling theory has been extensively applied to
inelastic scattering in liquids at $T\gtrsim T_{m}$, and this is the application for which we shall 
compare mode coupling and V-T theories.

Mode coupling theory works with the generalized Langevin equation for $F(q,t)$,
and expresses the memory function in terms of processes through which
density fluctuations decay \cite{BYbook, HMCDbook, BZbook}.  In the viscoelastic approximation, the
memory function decays with a $q$-dependent relaxation time \cite{BYbook, HMCDbook, BZbook}. This
approximation provides a good fit to the combined experimental data \cite{CRPRL74} and
MD data \cite{ARPRL74,ARPRA74} for the Brillouin peak dispersion curve in liquid Rb \cite{CLRPP38} (see also Fig.~9.2
of \cite{HMCDbook}). Going beyond the viscoelastic approximation,
Bosse et al.~\cite{BGLPRA17a, BGLPRA17b} constructed a self-consistent theory for the
longitudinal and transverse current fluctuation spectra, each expressed in
terms of relaxation kernels approximated by decay integrals which couple
the longitudinal and transverse excitations. This theory is in good overall
agreement with extensive neutron scattering data and MD calculations
for Ar near its triple point \cite{BGLPRA17b}. The theory was developed further by
Sj\"ogren \cite{Sjog80, Sjog80b}, who separated the memory function into a binary
collision part, approximated with a Gaussian ansatz, and a more collective
tail represented by a mode coupling term. For liquid Rb, this theory
gives an ``almost quantitative'' agreement with results from neutron scattering experiments
\cite{CRPRL74} and MD calculations \cite{ARPRL74,ARPRA74}. More recently, inelastic x-ray scattering 
measurements have been done for the
light alkali metals Li \cite{R&c002} and Na \cite{SBRSb,Naexp}. 
These data have been analyzed by mode coupling theory, and the resulting 
fits to $S(q,\omega)$ are excellent, both for the experimental data and for MD
calculations \cite{SBRSb,SRSVPRE02,SRSVPM02,SBRS,SBRSa}.

Recent results of V-T theory exhibit the following properties for the example of 
liquid sodium at melt. (a) The vibrational contribution alone, with no adjustable 
parameters, gives an excellent account of the Brillouin peak location at all $q$,
and gives the natural width of the Brillouin peak, which is at least half the
total width \cite{ARXIV05}. (b) The model described in Section~IV, with two
relaxation rates and a single multiplicative factor, can be made to fit $S(q,\omega)$~vs~$\omega$ 
for a wide range of $q$, with relaxation rates near the transit rate as theoretically 
predicted \cite{ARXIV05m}.  The result illustrates the important point of comparison 
between V-T and mode coupling theories: the two methods are based on different 
decompositions of the physical processes involved. While mode coupling theory 
analyzes $F(q,t)$ in terms of processes by which density fluctuations decay, V-T 
theory analyzes $F(q,t)$ in terms of the two contributions to the total liquid motion, 
vibrations and transits.  These two parts contribute to $S(q,\omega)$ as follows. (a) The
extended random valley vibrational modes have infinite lifetimes, and represent no
decay processes at all, yet they give the Brillouin peak its natural width. (b) In
zeroth order where anharmonicity is neglected, transits are responsible for all
genuine decay processes, and they contribute the entire width of the Rayleigh peak
and part of the width of the Brillouin peak.

\section{Summary and Discussion}

The configuration part of an equilibrium statistical mechanical average
can be thought of as an average over the many-particle potential energy surface.
In V-T theory this average is constructed from integrals over individual potential
energy valleys, where the entire contribution in the thermodynamic limit is
from random valleys. Formally exact formulas are given for the canonical
partition function in Eqs.~(\ref{eq2}) and (\ref{eq3}), and for the time correlation function
$\left<B(t) B^{*}(0)\right>$ in Eq.~(\ref{eq6}). For time-independent averages, which give the 
thermodynamic functions, the configuration contribution is the same for
every random valley, as in Eq.~(\ref{eq5b}).
 However, the same simplification does not hold for $\left<B(t) B^{*}(0)\right>$,
because the system may be in a different random valley at time $t$ than at
time $0$.

In a weighted integral over a random valley, the valley-valley
intersections give only a small contribution, because these intersections are 
present in a relatively small part of configuration space. Hence in each such 
integral, only a small error is made when the actual random valley is replaced 
by its harmonic extension to infinity, and the integral is extended
as well. We model the vibrational contribution to any equilibrium statistical
mechanical average by that average evaluated for a system moving in one extended 
random valley. This model has no adjustable parameters, and can
be evaluated for any temperature. The neglect of intervalley intersections,
and the incidental neglect of anharmonicity, can be corrected for when
necessary. The vibrational contribution to the partition function is
written in Eq.~(\ref{eq16}), and to a time correlation function in Eq.~(\ref{eq17}).

By way of example, the vibrational contribution is studied for the dynamic
response functions $F(q,t)$ and $S(q,\omega)$. The inelastic part in the one-mode
approximation, $S_{vib}^{(1)}(q,\omega)$, consists of independent scattering from each of the
complete set of normal vibrational modes. This scattering determines the location
and natural width of the Brillouin peak. Vibrational theory is in near
 perfect agreement with MD calculations, Fig.~1.
 
 In contrast to the thermodynamic functions, time correlation functions are 
 strongly influenced by transits. Transits cause the system trajectory to 
 move among random valleys, and in the dynamical variable $B(t)B^{*}(0)$, 
 each transit contributes to the decorrelation of $B(t)$ with $B^{*}(0)$. A simple
 approximation for $F(q,t)$, Eq.~(\ref{eq41}), preserves this effect in the atomic
 equilibrium positions, but neglects it in the atomic displacements. This 
 approximation yields two results: (a) while  with vibrational motion alone we
 have $F_{vib}(q,\infty)>0$,  when transits are present we have $F_{liq}(q,\infty)=0$,
 and (b) while with vibrational motion alone the Rayleigh peak has zero 
 width and the Brillouin peak has its nonzero natural width, the effect
 of transits is to broaden both peaks without shifting them.
 These transit properties are incorporated into a parametrized model, Eqs.~(\ref{eq42}-\ref{eq46}),
 which is capable of accounting for MD calculations of $S(q,\omega)$ in the 
 liquid Fig.~2.
 
 V-T theory rests on a new analysis of the many-body potential 
 energy surface underlying the motion of a monatomic liquid, namely
 the symmetry classification of valleys and the statistical dominance and microscopic
 uniformity of the random valleys. The same potential surface is then appropriate 
 for supercooled liquid and glass states. The glass corresponds
 to very small transit rate \cite{VT10}, vanishing on the timescale of dynamic response
 experiments, so property (a) of the preceding paragraph implies that $F(q,\infty)$ is
 positive for the glass. This result has been observed for real glasses  \cite{Mez89} and
 also for computer models of glasses \cite{deLeew88,deLeew90}. From V-T theory, the value of
 this positive constant is $F_{vib}(q,\infty)$ for an extended random valley. It has also 
 been shown previously that a vibrational analysis agrees with MD calculations
 for  $S(q,\omega)$ in a LJ glass \cite{S1,R&cPRL00}. The similar result for sodium
 is shown here in Fig.~1, where the vibrational theory is specifically for an extended
 random valley. Then, according to  V-T theory, this
 vibrational contribution has application to the liquid state, and leads to the
 result shown in Fig.~2.

 Finally, we have also compared and contrasted our results with those of the potential energy landscape (PEL)
  and mode coupling theory (MCT) approaches to liquid dynamics.  The various types of PEL theory seek to
   describe the statistical mechanics of liquids in terms of temperature-dependent properties of their potential
    energy surfaces, while MCT describes the liquid in terms of different treatments of the Langevin equation for
	 $F(q,t)$.  In contrast, our goal is to produce a theory with an accurate and tractable zeroth order Hamiltonian,
from which equilibrium and nonequilibrium quantities can be computed without adjustable parameters, 
and then make corrections to this Hamiltonian.  In so doing, we have predicted and verified several 
significant properties of the potential energy surface, such as the classification of valleys 
described in the previous paragraph,
 making the analysis of liquid dynamics considerably simpler.  
We have then identified two distinct corrections to the zeroth order Hamiltonian,
namely the boundary correction which accounts for valley-valley intersections, and 
the anharmonicity of the random valleys. Of these two, the boundary correction
 is by far the most important, and is not too difficult
to estimate. This helps address a significant problem faced recently
by PEL theories.  Further, some studies in the PEL program 
are revealing properties of the potential energy surface predicted by V-T theory independently of our efforts.  
In contrasting our treatment of $F(q,t)$ with MCT, we noted that our theories are based on different 
decompositions of the relevant physical processes:  MCT analyzes $F(q,t)$ in terms of the decay of density 
fluctuations, while V-T analysis is in terms of the two contributions to the actual motion of particles 
in a liquid, vibrations and transits.  We argue that the direct connection of this analysis to the motion 
of the particles, which is of universal applicability to all equilibrium and nonequilibrium quantities, 
provides a basis for a unified treatment of all processes in the liquid and supercooled states.

 There is much still to learn about the physical nature of transits. We have
 observed from the outset that transits must be ``correlation controlled,'' 
 and not just thermally activated \cite{VT5}. This is because, as mentioned
 above, the valley-valley intersections are present in a relatively 
 small part of configuration space, so that the transit rate is
 limited by the time required for the system to find a transit pathway.
 The highly correlated motion of the atoms involved in a transit 
 is revealed in MD calculations \cite{VT12}. Hence
 transits provide a new and challenging problem in the motion of many
 particle systems, and the solution of this problem will have a key role
 in the nonequilibrium properties of liquids, through their time
 correlation functions.

\begin{acknowledgments}  
 This work was supported by the US DOE through contract W-7405-ENG-36.
\end{acknowledgments}

\bibliography{genrefcorr}

\begin{thebibliography}{125}
\expandafter\ifx\csname natexlab\endcsname\relax\def\natexlab#1{#1}\fi
\expandafter\ifx\csname bibnamefont\endcsname\relax
  \def\bibnamefont#1{#1}\fi
\expandafter\ifx\csname bibfnamefont\endcsname\relax
  \def\bibfnamefont#1{#1}\fi
\expandafter\ifx\csname citenamefont\endcsname\relax
  \def\citenamefont#1{#1}\fi
\expandafter\ifx\csname url\endcsname\relax
  \def\url#1{\texttt{#1}}\fi
\expandafter\ifx\csname urlprefix\endcsname\relax\def\urlprefix{URL }\fi
\providecommand{\bibinfo}[2]{#2}
\providecommand{\eprint}[2][]{\url{#2}}

\bibitem[{\citenamefont{Frenkel}(1926)}]{E1}
\bibinfo{author}{\bibfnamefont{J.}~\bibnamefont{Frenkel}},
  \bibinfo{journal}{Z.\ Phys.} \textbf{\bibinfo{volume}{35}},
  \bibinfo{pages}{652} (\bibinfo{year}{1926}).

\bibitem[{\citenamefont{Frenkel}(1946)}]{E2}
\bibinfo{author}{\bibfnamefont{J.}~\bibnamefont{Frenkel}},
  \emph{\bibinfo{title}{Kinetic Theory of Liquids}}
  (\bibinfo{publisher}{Clarendon, Oxford}, \bibinfo{year}{1946}),
  \bibinfo{note}{chap.~III, Sec.~1}.

\bibitem[{\citenamefont{Singwi and Sj{\"o}lander}(1960)}]{E3}
\bibinfo{author}{\bibfnamefont{K.~S.} \bibnamefont{Singwi}} \bibnamefont{and}
  \bibinfo{author}{\bibfnamefont{A.}~\bibnamefont{Sj{\"o}lander}},
  \bibinfo{journal}{Phys.\ Rev.} \textbf{\bibinfo{volume}{119}},
  \bibinfo{pages}{863} (\bibinfo{year}{1960}).

\bibitem[{\citenamefont{Rahman et~al.}(1961)\citenamefont{Rahman, Singwi, and
  Sj{\"o}lander}}]{E4a}
\bibinfo{author}{\bibfnamefont{A.}~\bibnamefont{Rahman}},
  \bibinfo{author}{\bibfnamefont{K.~S.} \bibnamefont{Singwi}},
  \bibnamefont{and}
  \bibinfo{author}{\bibfnamefont{A.}~\bibnamefont{Sj{\"o}lander}},
  \bibinfo{journal}{Phys.\ Rev.} \textbf{\bibinfo{volume}{122}},
  \bibinfo{pages}{9} (\bibinfo{year}{1961}).

\bibitem[{\citenamefont{Rahman et~al.}(1962{\natexlab{a}})\citenamefont{Rahman,
  Singwi, and Sj{\"o}lander}}]{E4b}
\bibinfo{author}{\bibfnamefont{A.}~\bibnamefont{Rahman}},
  \bibinfo{author}{\bibfnamefont{K.~S.} \bibnamefont{Singwi}},
  \bibnamefont{and}
  \bibinfo{author}{\bibfnamefont{A.}~\bibnamefont{Sj{\"o}lander}},
  \bibinfo{journal}{Phys.\ Rev.} \textbf{\bibinfo{volume}{126}},
  \bibinfo{pages}{982} (\bibinfo{year}{1962}{\natexlab{a}}).

\bibitem[{\citenamefont{Rahman et~al.}(1962{\natexlab{b}})\citenamefont{Rahman,
  Singwi, and Sj{\"o}lander}}]{E4c}
\bibinfo{author}{\bibfnamefont{A.}~\bibnamefont{Rahman}},
  \bibinfo{author}{\bibfnamefont{K.~S.} \bibnamefont{Singwi}},
  \bibnamefont{and}
  \bibinfo{author}{\bibfnamefont{A.}~\bibnamefont{Sj{\"o}lander}},
  \bibinfo{journal}{Phys.\ Rev.} \textbf{\bibinfo{volume}{126}},
  \bibinfo{pages}{997} (\bibinfo{year}{1962}{\natexlab{b}}).

\bibitem[{\citenamefont{Damle et~al.}(1968)\citenamefont{Damle, Sj{\"o}lander,
  and Singwi}}]{E5}
\bibinfo{author}{\bibfnamefont{P.~S.} \bibnamefont{Damle}},
  \bibinfo{author}{\bibfnamefont{A.}~\bibnamefont{Sj{\"o}lander}},
  \bibnamefont{and} \bibinfo{author}{\bibfnamefont{K.~S.}
  \bibnamefont{Singwi}}, \bibinfo{journal}{Phys.\ Rev.}
  \textbf{\bibinfo{volume}{165}}, \bibinfo{pages}{277} (\bibinfo{year}{1968}).

\bibitem[{\citenamefont{Goldstein}(1969)}]{E6}
\bibinfo{author}{\bibfnamefont{M.}~\bibnamefont{Goldstein}},
  \bibinfo{journal}{J.\ Chem.\ Phys.} \textbf{\bibinfo{volume}{51}},
  \bibinfo{pages}{3728} (\bibinfo{year}{1969}).

\bibitem[{\citenamefont{Stillinger and Weber}(1982{\natexlab{a}})}]{E11}
\bibinfo{author}{\bibfnamefont{F.~H.} \bibnamefont{Stillinger}}
  \bibnamefont{and} \bibinfo{author}{\bibfnamefont{T.~A.} \bibnamefont{Weber}},
  \bibinfo{journal}{Phys.\ Rev.\ A} \textbf{\bibinfo{volume}{25}},
  \bibinfo{pages}{978} (\bibinfo{year}{1982}{\natexlab{a}}).

\bibitem[{\citenamefont{Stillinger and Weber}(1982{\natexlab{b}})}]{E12}
\bibinfo{author}{\bibfnamefont{F.~H.} \bibnamefont{Stillinger}}
  \bibnamefont{and} \bibinfo{author}{\bibfnamefont{T.~A.} \bibnamefont{Weber}},
  \bibinfo{journal}{Phys.\ Rev.\ A} \textbf{\bibinfo{volume}{28}},
  \bibinfo{pages}{2408} (\bibinfo{year}{1982}{\natexlab{b}}).

\bibitem[{\citenamefont{Stillinger and Weber}(1984)}]{E13}
\bibinfo{author}{\bibfnamefont{F.~H.} \bibnamefont{Stillinger}}
  \bibnamefont{and} \bibinfo{author}{\bibfnamefont{T.~A.} \bibnamefont{Weber}},
  \bibinfo{journal}{Science} \textbf{\bibinfo{volume}{225}},
  \bibinfo{pages}{983} (\bibinfo{year}{1984}).

\bibitem[{\citenamefont{Zwanzig}(1983)}]{E8}
\bibinfo{author}{\bibfnamefont{R.}~\bibnamefont{Zwanzig}},
  \bibinfo{journal}{J.\ Chem.\ Phys.} \textbf{\bibinfo{volume}{79}},
  \bibinfo{pages}{4507} (\bibinfo{year}{1983}).

\bibitem[{\citenamefont{Wallace}(1997{\natexlab{a}})}]{VT5}
\bibinfo{author}{\bibfnamefont{D.~C.} \bibnamefont{Wallace}},
  \bibinfo{journal}{Phys.\ Rev.\ E} \textbf{\bibinfo{volume}{56}},
  \bibinfo{pages}{4179} (\bibinfo{year}{1997}{\natexlab{a}}).

\bibitem[{\citenamefont{Wallace}(1991)}]{VT2}
\bibinfo{author}{\bibfnamefont{D.~C.} \bibnamefont{Wallace}},
  \bibinfo{journal}{Proc.\ R.\ Soc.\ Lond.\ A} \textbf{\bibinfo{volume}{433}},
  \bibinfo{pages}{631} (\bibinfo{year}{1991}).

\bibitem[{\citenamefont{Wallace}(1992)}]{VT3}
\bibinfo{author}{\bibfnamefont{D.~C.} \bibnamefont{Wallace}},
  \bibinfo{journal}{Proc.\ R.\ Soc.\ Lond.\ A} \textbf{\bibinfo{volume}{439}},
  \bibinfo{pages}{177} (\bibinfo{year}{1992}).

\bibitem[{\citenamefont{Wallace}(1997{\natexlab{b}})}]{Vt4}
\bibinfo{author}{\bibfnamefont{D.~C.} \bibnamefont{Wallace}},
  \bibinfo{journal}{Phys.\ Rev.\ E} \textbf{\bibinfo{volume}{56}},
  \bibinfo{pages}{1981} (\bibinfo{year}{1997}{\natexlab{b}}).

\bibitem[{\citenamefont{Wallace and Clements}(1999)}]{VT8}
\bibinfo{author}{\bibfnamefont{D.~C.} \bibnamefont{Wallace}} \bibnamefont{and}
  \bibinfo{author}{\bibfnamefont{B.~E.} \bibnamefont{Clements}},
  \bibinfo{journal}{Phys.\ Rev.\ E} \textbf{\bibinfo{volume}{59}},
  \bibinfo{pages}{2942} (\bibinfo{year}{1999}).

\bibitem[{\citenamefont{Clements and Wallace}(1999)}]{VT9}
\bibinfo{author}{\bibfnamefont{B.~E.} \bibnamefont{Clements}} \bibnamefont{and}
  \bibinfo{author}{\bibfnamefont{D.~C.} \bibnamefont{Wallace}},
  \bibinfo{journal}{Phys.\ Rev.\ E} \textbf{\bibinfo{volume}{59}},
  \bibinfo{pages}{2955} (\bibinfo{year}{1999}).

\bibitem[{\citenamefont{Wallace}(1998)}]{VT6}
\bibinfo{author}{\bibfnamefont{D.~C.} \bibnamefont{Wallace}},
  \bibinfo{journal}{Phys.\ Rev.\ E} \textbf{\bibinfo{volume}{57}},
  \bibinfo{pages}{1717} (\bibinfo{year}{1998}).

\bibitem[{\citenamefont{Chisolm and Wallace}(2004)}]{VT14}
\bibinfo{author}{\bibfnamefont{E.~D.} \bibnamefont{Chisolm}} \bibnamefont{and}
  \bibinfo{author}{\bibfnamefont{D.~C.} \bibnamefont{Wallace}},
  \bibinfo{journal}{Phys.\ Rev.\ E} \textbf{\bibinfo{volume}{69}},
  \bibinfo{pages}{031204} (\bibinfo{year}{2004}).

\bibitem[{\citenamefont{Wallace et~al.}(2001)\citenamefont{Wallace, Chisolm,
  and Clements}}]{VT12}
\bibinfo{author}{\bibfnamefont{D.~C.} \bibnamefont{Wallace}},
  \bibinfo{author}{\bibfnamefont{E.~D.} \bibnamefont{Chisolm}},
  \bibnamefont{and} \bibinfo{author}{\bibfnamefont{B.~E.}
  \bibnamefont{Clements}}, \bibinfo{journal}{Phys.\ Rev.\ E}
  \textbf{\bibinfo{volume}{64}}, \bibinfo{pages}{011205}
  (\bibinfo{year}{2001}).

\bibitem[{\citenamefont{Wallace}(1999)}]{VT10}
\bibinfo{author}{\bibfnamefont{D.~C.} \bibnamefont{Wallace}},
  \bibinfo{journal}{Phys.\ Rev.\ E} \textbf{\bibinfo{volume}{60}},
  \bibinfo{pages}{7049} (\bibinfo{year}{1999}).

\bibitem[{\citenamefont{Vollmayr et~al.}(1996)\citenamefont{Vollmayr, Kob, and
  Binder}}]{VKBJCP105}
\bibinfo{author}{\bibfnamefont{K.}~\bibnamefont{Vollmayr}},
  \bibinfo{author}{\bibfnamefont{W.}~\bibnamefont{Kob}}, \bibnamefont{and}
  \bibinfo{author}{\bibfnamefont{K.}~\bibnamefont{Binder}},
  \bibinfo{journal}{J.\ Chem.\ Phys.} \textbf{\bibinfo{volume}{105}},
  \bibinfo{pages}{4714} (\bibinfo{year}{1996}).

\bibitem[{\citenamefont{Chisolm et~al.}(2001)\citenamefont{Chisolm, Clements,
  and Wallace}}]{VT11}
\bibinfo{author}{\bibfnamefont{E.~D.} \bibnamefont{Chisolm}},
  \bibinfo{author}{\bibfnamefont{B.~E.} \bibnamefont{Clements}},
  \bibnamefont{and} \bibinfo{author}{\bibfnamefont{D.~C.}
  \bibnamefont{Wallace}}, \bibinfo{journal}{Phys.\ Rev.\ E}
  \textbf{\bibinfo{volume}{63}}, \bibinfo{pages}{031204}
  (\bibinfo{year}{2001}).

\bibitem[{\citenamefont{Chisolm and Wallace}(2001)}]{VT13}
\bibinfo{author}{\bibfnamefont{E.~D.} \bibnamefont{Chisolm}} \bibnamefont{and}
  \bibinfo{author}{\bibfnamefont{D.~C.} \bibnamefont{Wallace}},
  \bibinfo{journal}{J.\ Phys.:\ Condens. Matter} \textbf{\bibinfo{volume}{13}},
  \bibinfo{pages}{R739} (\bibinfo{year}{2001}).

\bibitem[{\citenamefont{Hansen and McDonald}(1986)}]{HMCDbook}
\bibinfo{author}{\bibfnamefont{J.~P.} \bibnamefont{Hansen}} \bibnamefont{and}
  \bibinfo{author}{\bibfnamefont{I.~R.} \bibnamefont{McDonald}},
  \emph{\bibinfo{title}{Theory of Simple Liquids}}
  (\bibinfo{publisher}{Academic, New York}, \bibinfo{year}{1986}),
  \bibinfo{edition}{2nd} ed.

\bibitem[{\citenamefont{Lovesey}(1984)}]{Lovbook}
\bibinfo{author}{\bibfnamefont{S.~W.} \bibnamefont{Lovesey}},
  \emph{\bibinfo{title}{Theory of Neutron Scattering from Condensed Matter}},
  vol.~\bibinfo{volume}{1} (\bibinfo{publisher}{Clarendon Press, Oxford},
  \bibinfo{year}{1984}).

\bibitem[{\citenamefont{Glyde}(1994)}]{Glybook}
\bibinfo{author}{\bibfnamefont{H.~R.} \bibnamefont{Glyde}},
  \emph{\bibinfo{title}{Excitations in Liquid and Solid Helium}}
  (\bibinfo{publisher}{Clarendon Press, Oxford}, \bibinfo{year}{1994}).

\bibitem[{\citenamefont{{De~Lorenzi-Venneri} and Wallace}()}]{ARXIV05m}
\bibinfo{author}{\bibfnamefont{G.}~\bibnamefont{{De~Lorenzi-Venneri}}}
  \bibnamefont{and} \bibinfo{author}{\bibfnamefont{D.~C.}
  \bibnamefont{Wallace}}, \bibinfo{howpublished}{arXiv: cond-mat/0509775, to be
  published in JCP}.

\bibitem[{\citenamefont{Wallace et~al.}()\citenamefont{Wallace,
  {De~Lorenzi-Venneri}, and Chisolm}}]{ARXIV05}
\bibinfo{author}{\bibfnamefont{D.~C.} \bibnamefont{Wallace}},
  \bibinfo{author}{\bibfnamefont{G.}~\bibnamefont{{De~Lorenzi-Venneri}}},
  \bibnamefont{and} \bibinfo{author}{\bibfnamefont{E.~D.}
  \bibnamefont{Chisolm}}, \bibinfo{howpublished}{arXiv: cond-mat/0506369}.

\bibitem[{\citenamefont{Wallace}(2002)}]{VT15}
\bibinfo{author}{\bibfnamefont{D.~C.} \bibnamefont{Wallace}},
  \emph{\bibinfo{title}{Statistical Physics of Crystals and Liquids}}
  (\bibinfo{publisher}{World Scientific, New Jersey}, \bibinfo{year}{2002}).

\bibitem[{\citenamefont{Rahman et~al.}(1976)\citenamefont{Rahman, Mandell, and
  McTague}}]{E7}
\bibinfo{author}{\bibfnamefont{A.}~\bibnamefont{Rahman}},
  \bibinfo{author}{\bibfnamefont{M.}~\bibnamefont{Mandell}}, \bibnamefont{and}
  \bibinfo{author}{\bibfnamefont{J.~P.} \bibnamefont{McTague}},
  \bibinfo{journal}{J.\ Chem.\ Phys.} \textbf{\bibinfo{volume}{64}},
  \bibinfo{pages}{1564} (\bibinfo{year}{1976}).

\bibitem[{\citenamefont{Seeley and Keyes}(1989)}]{AA}
\bibinfo{author}{\bibfnamefont{G.}~\bibnamefont{Seeley}} \bibnamefont{and}
  \bibinfo{author}{\bibfnamefont{T.}~\bibnamefont{Keyes}},
  \bibinfo{journal}{J.\ Chem.\ Phys.} \textbf{\bibinfo{volume}{91}},
  \bibinfo{pages}{5581} (\bibinfo{year}{1989}).

\bibitem[{\citenamefont{Madan et~al.}(1990)\citenamefont{Madan, Keyes, and
  Seeley}}]{AB}
\bibinfo{author}{\bibfnamefont{B.}~\bibnamefont{Madan}},
  \bibinfo{author}{\bibfnamefont{T.}~\bibnamefont{Keyes}}, \bibnamefont{and}
  \bibinfo{author}{\bibfnamefont{G.}~\bibnamefont{Seeley}},
  \bibinfo{journal}{J.\ Chem.\ Phys.} \textbf{\bibinfo{volume}{92}},
  \bibinfo{pages}{7565} (\bibinfo{year}{1990}).

\bibitem[{\citenamefont{Madan et~al.}(1991)\citenamefont{Madan, Keyes, and
  Seeley}}]{AC}
\bibinfo{author}{\bibfnamefont{B.}~\bibnamefont{Madan}},
  \bibinfo{author}{\bibfnamefont{T.}~\bibnamefont{Keyes}}, \bibnamefont{and}
  \bibinfo{author}{\bibfnamefont{G.}~\bibnamefont{Seeley}},
  \bibinfo{journal}{J.\ Chem.\ Phys.} \textbf{\bibinfo{volume}{94}},
  \bibinfo{pages}{6762} (\bibinfo{year}{1991}).

\bibitem[{\citenamefont{Madan and Keyes}(1993)}]{AE}
\bibinfo{author}{\bibfnamefont{B.}~\bibnamefont{Madan}} \bibnamefont{and}
  \bibinfo{author}{\bibfnamefont{T.}~\bibnamefont{Keyes}},
  \bibinfo{journal}{J.\ Chem.\ Phys.} \textbf{\bibinfo{volume}{98}},
  \bibinfo{pages}{3342} (\bibinfo{year}{1993}).

\bibitem[{\citenamefont{Keyes}(1994)}]{AG}
\bibinfo{author}{\bibfnamefont{T.}~\bibnamefont{Keyes}}, \bibinfo{journal}{J.\
  Chem.\ Phys.} \textbf{\bibinfo{volume}{101}}, \bibinfo{pages}{5081}
  (\bibinfo{year}{1994}).

\bibitem[{\citenamefont{Keyes}(1995)}]{AH}
\bibinfo{author}{\bibfnamefont{T.}~\bibnamefont{Keyes}}, \bibinfo{journal}{J.\
  Chem.\ Phys.} \textbf{\bibinfo{volume}{103}}, \bibinfo{pages}{9810}
  (\bibinfo{year}{1995}).

\bibitem[{\citenamefont{Bembenek and Laird}(1995)}]{CD}
\bibinfo{author}{\bibfnamefont{S.~D.} \bibnamefont{Bembenek}} \bibnamefont{and}
  \bibinfo{author}{\bibfnamefont{B.~B.} \bibnamefont{Laird}},
  \bibinfo{journal}{Phys.\ Rev.\ Lett.} \textbf{\bibinfo{volume}{74}},
  \bibinfo{pages}{936} (\bibinfo{year}{1995}).

\bibitem[{\citenamefont{Bembenek and Laird}(1996)}]{CE}
\bibinfo{author}{\bibfnamefont{S.~D.} \bibnamefont{Bembenek}} \bibnamefont{and}
  \bibinfo{author}{\bibfnamefont{B.~B.} \bibnamefont{Laird}},
  \bibinfo{journal}{J.\ Chem.\ Phys.} \textbf{\bibinfo{volume}{104}},
  \bibinfo{pages}{5199} (\bibinfo{year}{1996}).

\bibitem[{\citenamefont{Keyes et~al.}(1997)\citenamefont{Keyes, Vijayadamodar,
  and Zurcher}}]{AI}
\bibinfo{author}{\bibfnamefont{T.}~\bibnamefont{Keyes}},
  \bibinfo{author}{\bibfnamefont{G.~V.} \bibnamefont{Vijayadamodar}},
  \bibnamefont{and} \bibinfo{author}{\bibfnamefont{U.}~\bibnamefont{Zurcher}},
  \bibinfo{journal}{J.\ Chem.\ Phys.} \textbf{\bibinfo{volume}{106}},
  \bibinfo{pages}{4651} (\bibinfo{year}{1997}).

\bibitem[{\citenamefont{Li and Keyes}(1997)}]{AJ}
\bibinfo{author}{\bibfnamefont{W.-X.} \bibnamefont{Li}} \bibnamefont{and}
  \bibinfo{author}{\bibfnamefont{T.}~\bibnamefont{Keyes}},
  \bibinfo{journal}{J.\ Chem.\ Phys.} \textbf{\bibinfo{volume}{107}},
  \bibinfo{pages}{7275} (\bibinfo{year}{1997}).

\bibitem[{\citenamefont{Li et~al.}(1998)\citenamefont{Li, Keyes, and
  Sciortino}}]{AM}
\bibinfo{author}{\bibfnamefont{W.-X.} \bibnamefont{Li}},
  \bibinfo{author}{\bibfnamefont{T.}~\bibnamefont{Keyes}}, \bibnamefont{and}
  \bibinfo{author}{\bibfnamefont{F.}~\bibnamefont{Sciortino}},
  \bibinfo{journal}{J.\ Chem.\ Phys.} \textbf{\bibinfo{volume}{108}},
  \bibinfo{pages}{252} (\bibinfo{year}{1998}).

\bibitem[{\citenamefont{Li and Keyes}(1999)}]{AN}
\bibinfo{author}{\bibfnamefont{W.-X.} \bibnamefont{Li}} \bibnamefont{and}
  \bibinfo{author}{\bibfnamefont{T.}~\bibnamefont{Keyes}},
  \bibinfo{journal}{J.\ Chem.\ Phys.} \textbf{\bibinfo{volume}{111}},
  \bibinfo{pages}{5503} (\bibinfo{year}{1999}).

\bibitem[{\citenamefont{Buchner et~al.}(1992)\citenamefont{Buchner, Ladanyi,
  and Stratt}}]{BA}
\bibinfo{author}{\bibfnamefont{M.}~\bibnamefont{Buchner}},
  \bibinfo{author}{\bibfnamefont{B.~M.} \bibnamefont{Ladanyi}},
  \bibnamefont{and} \bibinfo{author}{\bibfnamefont{R.~M.}
  \bibnamefont{Stratt}}, \bibinfo{journal}{J.\ Chem.\ Phys.}
  \textbf{\bibinfo{volume}{97}}, \bibinfo{pages}{8522} (\bibinfo{year}{1992}).

\bibitem[{\citenamefont{Wu and Loring}(1992)}]{CA}
\bibinfo{author}{\bibfnamefont{T.-M.} \bibnamefont{Wu}} \bibnamefont{and}
  \bibinfo{author}{\bibfnamefont{R.~F.} \bibnamefont{Loring}},
  \bibinfo{journal}{J.\ Chem.\ Phys.} \textbf{\bibinfo{volume}{97}},
  \bibinfo{pages}{8568} (\bibinfo{year}{1992}).

\bibitem[{\citenamefont{Wu and Loring}(1993)}]{CB}
\bibinfo{author}{\bibfnamefont{T.-M.} \bibnamefont{Wu}} \bibnamefont{and}
  \bibinfo{author}{\bibfnamefont{R.~F.} \bibnamefont{Loring}},
  \bibinfo{journal}{J.\ Chem.\ Phys.} \textbf{\bibinfo{volume}{99}},
  \bibinfo{pages}{8936} (\bibinfo{year}{1993}).

\bibitem[{\citenamefont{Wan and Stratt}(1994)}]{BB}
\bibinfo{author}{\bibfnamefont{Y.}~\bibnamefont{Wan}} \bibnamefont{and}
  \bibinfo{author}{\bibfnamefont{R.~M.} \bibnamefont{Stratt}},
  \bibinfo{journal}{J.\ Chem.\ Phys.} \textbf{\bibinfo{volume}{100}},
  \bibinfo{pages}{5123} (\bibinfo{year}{1994}).

\bibitem[{\citenamefont{David and Stratt}(1998)}]{BF}
\bibinfo{author}{\bibfnamefont{E.~F.} \bibnamefont{David}} \bibnamefont{and}
  \bibinfo{author}{\bibfnamefont{R.~M.} \bibnamefont{Stratt}},
  \bibinfo{journal}{J.\ Chem.\ Phys.} \textbf{\bibinfo{volume}{109}},
  \bibinfo{pages}{1375} (\bibinfo{year}{1998}).

\bibitem[{\citenamefont{Kr{\"a}mer et~al.}(1998)\citenamefont{Kr{\"a}mer,
  Buchner, and Dorfm{\"u}ller}}]{CG}
\bibinfo{author}{\bibfnamefont{N.}~\bibnamefont{Kr{\"a}mer}},
  \bibinfo{author}{\bibfnamefont{M.}~\bibnamefont{Buchner}}, \bibnamefont{and}
  \bibinfo{author}{\bibfnamefont{T.}~\bibnamefont{Dorfm{\"u}ller}},
  \bibinfo{journal}{J.\ Chem.\ Phys.} \textbf{\bibinfo{volume}{109}},
  \bibinfo{pages}{1912} (\bibinfo{year}{1998}).

\bibitem[{\citenamefont{Ribeiro et~al.}(1998)\citenamefont{Ribeiro, Wilson, and
  Madden}}]{CF}
\bibinfo{author}{\bibfnamefont{M.~C.~C.} \bibnamefont{Ribeiro}},
  \bibinfo{author}{\bibfnamefont{M.}~\bibnamefont{Wilson}}, \bibnamefont{and}
  \bibinfo{author}{\bibfnamefont{P.~A.} \bibnamefont{Madden}},
  \bibinfo{journal}{J.\ Chem.\ Phys.} \textbf{\bibinfo{volume}{108}},
  \bibinfo{pages}{9027} (\bibinfo{year}{1998}).

\bibitem[{\citenamefont{Keyes and Fourkas}(2000)}]{AO}
\bibinfo{author}{\bibfnamefont{T.}~\bibnamefont{Keyes}} \bibnamefont{and}
  \bibinfo{author}{\bibfnamefont{J.~T.} \bibnamefont{Fourkas}},
  \bibinfo{journal}{J.\ Chem.\ Phys.} \textbf{\bibinfo{volume}{112}},
  \bibinfo{pages}{287} (\bibinfo{year}{2000}).

\bibitem[{\citenamefont{Vallauri and Bermejo}(1995)}]{CC}
\bibinfo{author}{\bibfnamefont{R.}~\bibnamefont{Vallauri}} \bibnamefont{and}
  \bibinfo{author}{\bibfnamefont{F.~J.} \bibnamefont{Bermejo}},
  \bibinfo{journal}{Phys.\ Rev.\ E} \textbf{\bibinfo{volume}{51}},
  \bibinfo{pages}{2654} (\bibinfo{year}{1995}).

\bibitem[{\citenamefont{Wu and Tsay}(1996)}]{CH}
\bibinfo{author}{\bibfnamefont{T.-M.} \bibnamefont{Wu}} \bibnamefont{and}
  \bibinfo{author}{\bibfnamefont{S.-F.} \bibnamefont{Tsay}},
  \bibinfo{journal}{J.\ Chem.\ Phys.} \textbf{\bibinfo{volume}{105}},
  \bibinfo{pages}{9281} (\bibinfo{year}{1996}).

\bibitem[{\citenamefont{Wu and Tsay}(1998)}]{CI}
\bibinfo{author}{\bibfnamefont{T.-M.} \bibnamefont{Wu}} \bibnamefont{and}
  \bibinfo{author}{\bibfnamefont{S.-F.} \bibnamefont{Tsay}},
  \bibinfo{journal}{Phys.\ Rev.\ B} \textbf{\bibinfo{volume}{58}},
  \bibinfo{pages}{27} (\bibinfo{year}{1998}).

\bibitem[{\citenamefont{Wu et~al.}(1998)\citenamefont{Wu, Ma, and Tsay}}]{CJ}
\bibinfo{author}{\bibfnamefont{T.-M.} \bibnamefont{Wu}},
  \bibinfo{author}{\bibfnamefont{W.-J.} \bibnamefont{Ma}}, \bibnamefont{and}
  \bibinfo{author}{\bibfnamefont{S.-F.} \bibnamefont{Tsay}},
  \bibinfo{journal}{Physica \ A} \textbf{\bibinfo{volume}{254}},
  \bibinfo{pages}{257} (\bibinfo{year}{1998}).

\bibitem[{\citenamefont{Wu et~al.}(2000)\citenamefont{Wu, Ma, and Chang}}]{CK}
\bibinfo{author}{\bibfnamefont{T.-M.} \bibnamefont{Wu}},
  \bibinfo{author}{\bibfnamefont{W.-J.} \bibnamefont{Ma}}, \bibnamefont{and}
  \bibinfo{author}{\bibfnamefont{S.-L.} \bibnamefont{Chang}},
  \bibinfo{journal}{J.\ Chem.\ Phys.} \textbf{\bibinfo{volume}{113}},
  \bibinfo{pages}{274} (\bibinfo{year}{2000}).

\bibitem[{\citenamefont{Cho et~al.}(1994)\citenamefont{Cho, Fleming, Saito,
  Ohmine, and Stratt}}]{BD}
\bibinfo{author}{\bibfnamefont{M.}~\bibnamefont{Cho}},
  \bibinfo{author}{\bibfnamefont{G.~R.} \bibnamefont{Fleming}},
  \bibinfo{author}{\bibfnamefont{S.}~\bibnamefont{Saito}},
  \bibinfo{author}{\bibfnamefont{S.}~\bibnamefont{Ohmine}}, \bibnamefont{and}
  \bibinfo{author}{\bibfnamefont{R.~M.} \bibnamefont{Stratt}},
  \bibinfo{journal}{J.\ Chem.\ Phys.} \textbf{\bibinfo{volume}{100}},
  \bibinfo{pages}{6672} (\bibinfo{year}{1994}).

\bibitem[{\citenamefont{{La Nave} et~al.}(2000)\citenamefont{{La Nave}, Scala,
  Starr, Sciortino, and Stanley}}]{HF}
\bibinfo{author}{\bibfnamefont{E.}~\bibnamefont{{La Nave}}},
  \bibinfo{author}{\bibfnamefont{A.}~\bibnamefont{Scala}},
  \bibinfo{author}{\bibfnamefont{F.~W.} \bibnamefont{Starr}},
  \bibinfo{author}{\bibfnamefont{F.}~\bibnamefont{Sciortino}},
  \bibnamefont{and} \bibinfo{author}{\bibfnamefont{H.~E.}
  \bibnamefont{Stanley}}, \bibinfo{journal}{Phys.\ Rev.\ Lett.}
  \textbf{\bibinfo{volume}{84}}, \bibinfo{pages}{4605} (\bibinfo{year}{2000}).

\bibitem[{\citenamefont{Starr et~al.}(2001)\citenamefont{Starr, Sastry, {La
  Nave}, Scala, Stanley, and Sciortino}}]{HG}
\bibinfo{author}{\bibfnamefont{F.~W.} \bibnamefont{Starr}},
  \bibinfo{author}{\bibfnamefont{S.}~\bibnamefont{Sastry}},
  \bibinfo{author}{\bibfnamefont{E.}~\bibnamefont{{La Nave}}},
  \bibinfo{author}{\bibfnamefont{A.}~\bibnamefont{Scala}},
  \bibinfo{author}{\bibfnamefont{H.~E.} \bibnamefont{Stanley}},
  \bibnamefont{and}
  \bibinfo{author}{\bibfnamefont{F.}~\bibnamefont{Sciortino}},
  \bibinfo{journal}{Phys.\ Rev.\ E} \textbf{\bibinfo{volume}{63}},
  \bibinfo{pages}{041201} (\bibinfo{year}{2001}).

\bibitem[{\citenamefont{{La Nave} et~al.}(2001)\citenamefont{{La Nave}, Scala,
  Starr, Stanley, and Sciortino}}]{HH}
\bibinfo{author}{\bibfnamefont{E.}~\bibnamefont{{La Nave}}},
  \bibinfo{author}{\bibfnamefont{A.}~\bibnamefont{Scala}},
  \bibinfo{author}{\bibfnamefont{F.~W.} \bibnamefont{Starr}},
  \bibinfo{author}{\bibfnamefont{H.~E.} \bibnamefont{Stanley}},
  \bibnamefont{and}
  \bibinfo{author}{\bibfnamefont{F.}~\bibnamefont{Sciortino}},
  \bibinfo{journal}{Phys.\ Rev.\ E} \textbf{\bibinfo{volume}{64}},
  \bibinfo{pages}{036102} (\bibinfo{year}{2001}).

\bibitem[{\citenamefont{Pohorille et~al.}(1987)\citenamefont{Pohorille, Pratt,
  {La~Violette}, and {MacElroy}}}]{DA}
\bibinfo{author}{\bibfnamefont{A.}~\bibnamefont{Pohorille}},
  \bibinfo{author}{\bibfnamefont{L.~R.} \bibnamefont{Pratt}},
  \bibinfo{author}{\bibfnamefont{R.~A.} \bibnamefont{{La~Violette}}},
  \bibnamefont{and} \bibinfo{author}{\bibfnamefont{R.~D.}
  \bibnamefont{{MacElroy}}}, \bibinfo{journal}{J.\ Chem.\ Phys.}
  \textbf{\bibinfo{volume}{87}}, \bibinfo{pages}{6070} (\bibinfo{year}{1987}).

\bibitem[{\citenamefont{Ohmine and Tanaka}(1993)}]{DB}
\bibinfo{author}{\bibfnamefont{I.}~\bibnamefont{Ohmine}} \bibnamefont{and}
  \bibinfo{author}{\bibfnamefont{H.}~\bibnamefont{Tanaka}},
  \bibinfo{journal}{Chem.\ Rev.} \textbf{\bibinfo{volume}{93}},
  \bibinfo{pages}{2545} (\bibinfo{year}{1993}).

\bibitem[{\citenamefont{Cao and Voth}(1995{\natexlab{a}})}]{DC}
\bibinfo{author}{\bibfnamefont{J.}~\bibnamefont{Cao}} \bibnamefont{and}
  \bibinfo{author}{\bibfnamefont{G.~A.} \bibnamefont{Voth}},
  \bibinfo{journal}{J.\ Chem.\ Phys.} \textbf{\bibinfo{volume}{102}},
  \bibinfo{pages}{3337} (\bibinfo{year}{1995}{\natexlab{a}}).

\bibitem[{\citenamefont{Cao and Voth}(1995{\natexlab{b}})}]{DD}
\bibinfo{author}{\bibfnamefont{J.}~\bibnamefont{Cao}} \bibnamefont{and}
  \bibinfo{author}{\bibfnamefont{G.~A.} \bibnamefont{Voth}},
  \bibinfo{journal}{J.\ Chem.\ Phys.} \textbf{\bibinfo{volume}{103}},
  \bibinfo{pages}{4211} (\bibinfo{year}{1995}{\natexlab{b}}).

\bibitem[{\citenamefont{Gezelter et~al.}(1997)\citenamefont{Gezelter, Rabani,
  and Berne}}]{DEa}
\bibinfo{author}{\bibfnamefont{J.~D.} \bibnamefont{Gezelter}},
  \bibinfo{author}{\bibfnamefont{E.}~\bibnamefont{Rabani}}, \bibnamefont{and}
  \bibinfo{author}{\bibfnamefont{B.~J.} \bibnamefont{Berne}},
  \bibinfo{journal}{J.\ Chem.\ Phys.} \textbf{\bibinfo{volume}{107}},
  \bibinfo{pages}{4618} (\bibinfo{year}{1997}).

\bibitem[{\citenamefont{Keyes et~al.}(1998)\citenamefont{Keyes, Li, and
  Zurcher}}]{DEb}
\bibinfo{author}{\bibfnamefont{T.}~\bibnamefont{Keyes}},
  \bibinfo{author}{\bibfnamefont{W.-X.} \bibnamefont{Li}}, \bibnamefont{and}
  \bibinfo{author}{\bibfnamefont{U.}~\bibnamefont{Zurcher}},
  \bibinfo{journal}{J.\ Chem.\ Phys.} \textbf{\bibinfo{volume}{109}},
  \bibinfo{pages}{4693} (\bibinfo{year}{1998}).

\bibitem[{\citenamefont{Gezelter et~al.}(1998)\citenamefont{Gezelter, Rabani,
  and Berne}}]{DEc}
\bibinfo{author}{\bibfnamefont{J.~D.} \bibnamefont{Gezelter}},
  \bibinfo{author}{\bibfnamefont{E.}~\bibnamefont{Rabani}}, \bibnamefont{and}
  \bibinfo{author}{\bibfnamefont{B.~J.} \bibnamefont{Berne}},
  \bibinfo{journal}{J.\ Chem.\ Phys.} \textbf{\bibinfo{volume}{109}},
  \bibinfo{pages}{4695} (\bibinfo{year}{1998}).

\bibitem[{\citenamefont{Rabani et~al.}(1997)\citenamefont{Rabani, Gezelter, and
  Berne}}]{DF}
\bibinfo{author}{\bibfnamefont{E.}~\bibnamefont{Rabani}},
  \bibinfo{author}{\bibfnamefont{J.~D.} \bibnamefont{Gezelter}},
  \bibnamefont{and} \bibinfo{author}{\bibfnamefont{B.~J.} \bibnamefont{Berne}},
  \bibinfo{journal}{J.\ Chem.\ Phys.} \textbf{\bibinfo{volume}{107}},
  \bibinfo{pages}{6867} (\bibinfo{year}{1997}).

\bibitem[{\citenamefont{Gezelter et~al.}(1999)\citenamefont{Gezelter, Rabani,
  and Berne}}]{DG}
\bibinfo{author}{\bibfnamefont{J.~D.} \bibnamefont{Gezelter}},
  \bibinfo{author}{\bibfnamefont{E.}~\bibnamefont{Rabani}}, \bibnamefont{and}
  \bibinfo{author}{\bibfnamefont{B.~J.} \bibnamefont{Berne}},
  \bibinfo{journal}{J.\ Chem.\ Phys.} \textbf{\bibinfo{volume}{110}},
  \bibinfo{pages}{3444} (\bibinfo{year}{1999}).

\bibitem[{\citenamefont{Kob and Andersen}(1994)}]{LA}
\bibinfo{author}{\bibfnamefont{W.}~\bibnamefont{Kob}} \bibnamefont{and}
  \bibinfo{author}{\bibfnamefont{H.~C.} \bibnamefont{Andersen}},
  \bibinfo{journal}{Phys.\ Rev.\ Lett.} \textbf{\bibinfo{volume}{73}},
  \bibinfo{pages}{1376} (\bibinfo{year}{1994}).

\bibitem[{\citenamefont{Sciortino et~al.}(2000)\citenamefont{Sciortino, Kob,
  and Tartaglia}}]{FC}
\bibinfo{author}{\bibfnamefont{F.}~\bibnamefont{Sciortino}},
  \bibinfo{author}{\bibfnamefont{W.}~\bibnamefont{Kob}}, \bibnamefont{and}
  \bibinfo{author}{\bibfnamefont{P.}~\bibnamefont{Tartaglia}},
  \bibinfo{journal}{J.\ Phys.:\ Condens. Matter} \textbf{\bibinfo{volume}{12}},
  \bibinfo{pages}{6525} (\bibinfo{year}{2000}).

\bibitem[{\citenamefont{Sciortino et~al.}(1999)\citenamefont{Sciortino, Kob,
  and Tartaglia}}]{FA}
\bibinfo{author}{\bibfnamefont{F.}~\bibnamefont{Sciortino}},
  \bibinfo{author}{\bibfnamefont{W.}~\bibnamefont{Kob}}, \bibnamefont{and}
  \bibinfo{author}{\bibfnamefont{P.}~\bibnamefont{Tartaglia}},
  \bibinfo{journal}{Phys.\ Rev.\ Lett.} \textbf{\bibinfo{volume}{83}},
  \bibinfo{pages}{3214} (\bibinfo{year}{1999}).

\bibitem[{\citenamefont{Sampoli et~al.}(2003)\citenamefont{Sampoli, Benassi,
  Eramo, Angelani, and Ruocco}}]{FB}
\bibinfo{author}{\bibfnamefont{M.}~\bibnamefont{Sampoli}},
  \bibinfo{author}{\bibfnamefont{P.}~\bibnamefont{Benassi}},
  \bibinfo{author}{\bibfnamefont{R.}~\bibnamefont{Eramo}},
  \bibinfo{author}{\bibfnamefont{L.}~\bibnamefont{Angelani}}, \bibnamefont{and}
  \bibinfo{author}{\bibfnamefont{G.}~\bibnamefont{Ruocco}},
  \bibinfo{journal}{J.\ Phys.:\ Condens. Matter} \textbf{\bibinfo{volume}{15}},
  \bibinfo{pages}{51227} (\bibinfo{year}{2003}).

\bibitem[{\citenamefont{B{\"u}chner and Heuer}(1999)}]{FE}
\bibinfo{author}{\bibfnamefont{S.}~\bibnamefont{B{\"u}chner}} \bibnamefont{and}
  \bibinfo{author}{\bibfnamefont{A.}~\bibnamefont{Heuer}},
  \bibinfo{journal}{Phys.\ Rev.\ E} \textbf{\bibinfo{volume}{60}},
  \bibinfo{pages}{6507} (\bibinfo{year}{1999}).

\bibitem[{\citenamefont{Sastry}(2000)}]{FF}
\bibinfo{author}{\bibfnamefont{S.}~\bibnamefont{Sastry}}, \bibinfo{journal}{J.\
  Phys.:\ Condens. Matter} \textbf{\bibinfo{volume}{12}}, \bibinfo{pages}{6515}
  (\bibinfo{year}{2000}).

\bibitem[{\citenamefont{Angelani et~al.}(2000)\citenamefont{Angelani,
  {Di~Leonardo}, Ruocco, Scala, and Sciortino}}]{IA}
\bibinfo{author}{\bibfnamefont{L.}~\bibnamefont{Angelani}},
  \bibinfo{author}{\bibfnamefont{R.}~\bibnamefont{{Di~Leonardo}}},
  \bibinfo{author}{\bibfnamefont{G.}~\bibnamefont{Ruocco}},
  \bibinfo{author}{\bibfnamefont{A.}~\bibnamefont{Scala}}, \bibnamefont{and}
  \bibinfo{author}{\bibfnamefont{F.}~\bibnamefont{Sciortino}},
  \bibinfo{journal}{Phys.\ Rev.\ Lett.} \textbf{\bibinfo{volume}{85}},
  \bibinfo{pages}{5356} (\bibinfo{year}{2000}).

\bibitem[{\citenamefont{Sastry et~al.}(1998)\citenamefont{Sastry, Debenedetti,
  and Stillinger}}]{JC}
\bibinfo{author}{\bibfnamefont{S.}~\bibnamefont{Sastry}},
  \bibinfo{author}{\bibfnamefont{P.~G.} \bibnamefont{Debenedetti}},
  \bibnamefont{and} \bibinfo{author}{\bibfnamefont{F.~H.}
  \bibnamefont{Stillinger}}, \bibinfo{journal}{Nature}
  \textbf{\bibinfo{volume}{393}}, \bibinfo{pages}{554} (\bibinfo{year}{1998}).

\bibitem[{\citenamefont{Schr{\o}der et~al.}(2000)\citenamefont{Schr{\o}der,
  Sastry, Dyre, and Glotzer}}]{JH}
\bibinfo{author}{\bibfnamefont{T.~B.} \bibnamefont{Schr{\o}der}},
  \bibinfo{author}{\bibfnamefont{S.}~\bibnamefont{Sastry}},
  \bibinfo{author}{\bibfnamefont{J.~C.} \bibnamefont{Dyre}}, \bibnamefont{and}
  \bibinfo{author}{\bibfnamefont{S.~C.} \bibnamefont{Glotzer}},
  \bibinfo{journal}{J.\ Chem.\ Phys.} \textbf{\bibinfo{volume}{112}},
  \bibinfo{pages}{9834} (\bibinfo{year}{2000}).

\bibitem[{\citenamefont{Angelani et~al.}(2002)\citenamefont{Angelani,
  {Di~Leonardo}, Ruocco, Scala, and Sciortino}}]{IC}
\bibinfo{author}{\bibfnamefont{L.}~\bibnamefont{Angelani}},
  \bibinfo{author}{\bibfnamefont{R.}~\bibnamefont{{Di~Leonardo}}},
  \bibinfo{author}{\bibfnamefont{G.}~\bibnamefont{Ruocco}},
  \bibinfo{author}{\bibfnamefont{A.}~\bibnamefont{Scala}}, \bibnamefont{and}
  \bibinfo{author}{\bibfnamefont{F.}~\bibnamefont{Sciortino}},
  \bibinfo{journal}{J.\ Chem.\ Phys.} \textbf{\bibinfo{volume}{116}},
  \bibinfo{pages}{10297} (\bibinfo{year}{2002}).

\bibitem[{\citenamefont{Angelani et~al.}(2003)\citenamefont{Angelani, Ruocco,
  Sampoli, and Sciortino}}]{ID}
\bibinfo{author}{\bibfnamefont{L.}~\bibnamefont{Angelani}},
  \bibinfo{author}{\bibfnamefont{G.}~\bibnamefont{Ruocco}},
  \bibinfo{author}{\bibfnamefont{M.}~\bibnamefont{Sampoli}}, \bibnamefont{and}
  \bibinfo{author}{\bibfnamefont{F.}~\bibnamefont{Sciortino}},
  \bibinfo{journal}{J.\ Chem.\ Phys.} \textbf{\bibinfo{volume}{119}},
  \bibinfo{pages}{2120} (\bibinfo{year}{2003}).

\bibitem[{\citenamefont{Scala et~al.}(2000)\citenamefont{Scala, Starr, {La
  Nave}, Sciortino, and Stanley}}]{HA}
\bibinfo{author}{\bibfnamefont{A.}~\bibnamefont{Scala}},
  \bibinfo{author}{\bibfnamefont{F.~W.} \bibnamefont{Starr}},
  \bibinfo{author}{\bibfnamefont{E.}~\bibnamefont{{La Nave}}},
  \bibinfo{author}{\bibfnamefont{F.}~\bibnamefont{Sciortino}},
  \bibnamefont{and} \bibinfo{author}{\bibfnamefont{H.~E.}
  \bibnamefont{Stanley}}, \bibinfo{journal}{Nature}
  \textbf{\bibinfo{volume}{406}}, \bibinfo{pages}{166} (\bibinfo{year}{2000}).

\bibitem[{\citenamefont{Sastry}(2001)}]{HB}
\bibinfo{author}{\bibfnamefont{S.}~\bibnamefont{Sastry}},
  \bibinfo{journal}{Nature} \textbf{\bibinfo{volume}{409}},
  \bibinfo{pages}{164} (\bibinfo{year}{2001}).

\bibitem[{\citenamefont{Saika-Voivod et~al.}(2001)\citenamefont{Saika-Voivod,
  Poole, and Sciortino}}]{HC}
\bibinfo{author}{\bibfnamefont{I.}~\bibnamefont{Saika-Voivod}},
  \bibinfo{author}{\bibfnamefont{P.~H.} \bibnamefont{Poole}}, \bibnamefont{and}
  \bibinfo{author}{\bibfnamefont{F.}~\bibnamefont{Sciortino}},
  \bibinfo{journal}{Nature} \textbf{\bibinfo{volume}{412}},
  \bibinfo{pages}{514} (\bibinfo{year}{2001}).

\bibitem[{\citenamefont{Debenedetti et~al.}(2003)\citenamefont{Debenedetti,
  Stillinger, and Shell}}]{FD}
\bibinfo{author}{\bibfnamefont{P.~G.} \bibnamefont{Debenedetti}},
  \bibinfo{author}{\bibfnamefont{F.~H.} \bibnamefont{Stillinger}},
  \bibnamefont{and} \bibinfo{author}{\bibfnamefont{M.~S.} \bibnamefont{Shell}},
  \bibinfo{journal}{J.\ Phys.\ Chem.\ B} \textbf{\bibinfo{volume}{107}},
  \bibinfo{pages}{14434} (\bibinfo{year}{2003}).

\bibitem[{\citenamefont{Heuer and B{\"u}chner}(2000)}]{S4}
\bibinfo{author}{\bibfnamefont{A.}~\bibnamefont{Heuer}} \bibnamefont{and}
  \bibinfo{author}{\bibfnamefont{S.}~\bibnamefont{B{\"u}chner}},
  \bibinfo{journal}{J.\ Phys.:\ Condens. Matter} \textbf{\bibinfo{volume}{12}},
  \bibinfo{pages}{6535} (\bibinfo{year}{2000}).

\bibitem[{\citenamefont{Adams and Gibbs}(1965)}]{E9}
\bibinfo{author}{\bibfnamefont{G.}~\bibnamefont{Adams}} \bibnamefont{and}
  \bibinfo{author}{\bibfnamefont{J.~H.} \bibnamefont{Gibbs}},
  \bibinfo{journal}{J.\ Chem.\ Phys.} \textbf{\bibinfo{volume}{43}},
  \bibinfo{pages}{139} (\bibinfo{year}{1965}).

\bibitem[{\citenamefont{Richert and Angell}(1998)}]{HD}
\bibinfo{author}{\bibfnamefont{R.}~\bibnamefont{Richert}} \bibnamefont{and}
  \bibinfo{author}{\bibfnamefont{C.~A.} \bibnamefont{Angell}},
  \bibinfo{journal}{J.\ Chem.\ Phys.} \textbf{\bibinfo{volume}{108}},
  \bibinfo{pages}{9016} (\bibinfo{year}{1998}).

\bibitem[{\citenamefont{Shell and Debenedetti}(2004)}]{HE}
\bibinfo{author}{\bibfnamefont{M.~S.} \bibnamefont{Shell}} \bibnamefont{and}
  \bibinfo{author}{\bibfnamefont{P.~G.} \bibnamefont{Debenedetti}},
  \bibinfo{journal}{Phys.\ Rev.\ E} \textbf{\bibinfo{volume}{69}},
  \bibinfo{pages}{051102} (\bibinfo{year}{2004}).

\bibitem[{\citenamefont{Mossa et~al.}(2002{\natexlab{a}})\citenamefont{Mossa,
  Ruocco, Sciortino, and Tartaglia}}]{JD}
\bibinfo{author}{\bibfnamefont{S.}~\bibnamefont{Mossa}},
  \bibinfo{author}{\bibfnamefont{G.}~\bibnamefont{Ruocco}},
  \bibinfo{author}{\bibfnamefont{F.}~\bibnamefont{Sciortino}},
  \bibnamefont{and}
  \bibinfo{author}{\bibfnamefont{P.}~\bibnamefont{Tartaglia}},
  \bibinfo{journal}{Phil.\ Mag.\ B} \textbf{\bibinfo{volume}{82}},
  \bibinfo{pages}{695} (\bibinfo{year}{2002}{\natexlab{a}}).

\bibitem[{\citenamefont{Sciortino and Tartaglia}(2001)}]{JG}
\bibinfo{author}{\bibfnamefont{F.}~\bibnamefont{Sciortino}} \bibnamefont{and}
  \bibinfo{author}{\bibfnamefont{P.}~\bibnamefont{Tartaglia}},
  \bibinfo{journal}{J.\ Phys.:\ Condens. Matter} \textbf{\bibinfo{volume}{13}},
  \bibinfo{pages}{9127} (\bibinfo{year}{2001}).

\bibitem[{\citenamefont{Donati et~al.}(2000)\citenamefont{Donati, Sciortino,
  and Tartaglia}}]{JI}
\bibinfo{author}{\bibfnamefont{C.}~\bibnamefont{Donati}},
  \bibinfo{author}{\bibfnamefont{F.}~\bibnamefont{Sciortino}},
  \bibnamefont{and}
  \bibinfo{author}{\bibfnamefont{P.}~\bibnamefont{Tartaglia}},
  \bibinfo{journal}{Phys.\ Rev.\ Lett.} \textbf{\bibinfo{volume}{85}},
  \bibinfo{pages}{1464} (\bibinfo{year}{2000}).

\bibitem[{\citenamefont{Kob et~al.}(2000)\citenamefont{Kob, Sciortino, and
  Tartaglia}}]{JK}
\bibinfo{author}{\bibfnamefont{W.}~\bibnamefont{Kob}},
  \bibinfo{author}{\bibfnamefont{F.}~\bibnamefont{Sciortino}},
  \bibnamefont{and}
  \bibinfo{author}{\bibfnamefont{P.}~\bibnamefont{Tartaglia}},
  \bibinfo{journal}{Europhys.\ Lett.} \textbf{\bibinfo{volume}{49}},
  \bibinfo{pages}{590} (\bibinfo{year}{2000}).

\bibitem[{\citenamefont{Sciortino}(2005)}]{S3}
\bibinfo{author}{\bibfnamefont{F.}~\bibnamefont{Sciortino}},
  \bibinfo{journal}{J.\ Stat. \ Mech.:\ Theory and Experiment}
  \textbf{\bibinfo{volume}{2}}, \bibinfo{pages}{P05015} (\bibinfo{year}{2005}).

\bibitem[{\citenamefont{Stratt}(1995)}]{BE}
\bibinfo{author}{\bibfnamefont{R.~M.} \bibnamefont{Stratt}},
  \bibinfo{journal}{Acc.\ Chem.\ Res.} \textbf{\bibinfo{volume}{28}},
  \bibinfo{pages}{201} (\bibinfo{year}{1995}).

\bibitem[{\citenamefont{Mossa et~al.}(2002{\natexlab{b}})\citenamefont{Mossa,
  {La Nave}, Stanley, Donati, Sciortino, Kob, and Tartaglia}}]{FG}
\bibinfo{author}{\bibfnamefont{S.}~\bibnamefont{Mossa}},
  \bibinfo{author}{\bibfnamefont{E.}~\bibnamefont{{La Nave}}},
  \bibinfo{author}{\bibfnamefont{H.~E.} \bibnamefont{Stanley}},
  \bibinfo{author}{\bibfnamefont{C.}~\bibnamefont{Donati}},
  \bibinfo{author}{\bibfnamefont{F.}~\bibnamefont{Sciortino}},
  \bibinfo{author}{\bibfnamefont{W.}~\bibnamefont{Kob}}, \bibnamefont{and}
  \bibinfo{author}{\bibfnamefont{P.}~\bibnamefont{Tartaglia}},
  \bibinfo{journal}{Phys.\ Rev.\ E} \textbf{\bibinfo{volume}{65}},
  \bibinfo{pages}{041205} (\bibinfo{year}{2002}{\natexlab{b}}).

\bibitem[{\citenamefont{{La Nave} et~al.}(2003)\citenamefont{{La Nave},
  Sciortino, Kob, Tartaglia, {De Michele}, and Mossa}}]{FH}
\bibinfo{author}{\bibfnamefont{E.}~\bibnamefont{{La Nave}}},
  \bibinfo{author}{\bibfnamefont{F.}~\bibnamefont{Sciortino}},
  \bibinfo{author}{\bibfnamefont{W.}~\bibnamefont{Kob}},
  \bibinfo{author}{\bibfnamefont{P.}~\bibnamefont{Tartaglia}},
  \bibinfo{author}{\bibfnamefont{C.}~\bibnamefont{{De Michele}}},
  \bibnamefont{and} \bibinfo{author}{\bibfnamefont{S.}~\bibnamefont{Mossa}},
  \bibinfo{journal}{J.\ Phys.:\ Condens. Matter} \textbf{\bibinfo{volume}{15}},
  \bibinfo{pages}{51085} (\bibinfo{year}{2003}).

\bibitem[{\citenamefont{Broderix et~al.}(2000)\citenamefont{Broderix,
  Bhattacharya, Cavagna, Zippelius, and Giardina}}]{IB}
\bibinfo{author}{\bibfnamefont{K.}~\bibnamefont{Broderix}},
  \bibinfo{author}{\bibfnamefont{K.~K.} \bibnamefont{Bhattacharya}},
  \bibinfo{author}{\bibfnamefont{A.}~\bibnamefont{Cavagna}},
  \bibinfo{author}{\bibfnamefont{A.}~\bibnamefont{Zippelius}},
  \bibnamefont{and} \bibinfo{author}{\bibfnamefont{I.}~\bibnamefont{Giardina}},
  \bibinfo{journal}{Phys.\ Rev.\ Lett.} \textbf{\bibinfo{volume}{85}},
  \bibinfo{pages}{5360} (\bibinfo{year}{2000}).

\bibitem[{\citenamefont{Boon and Yip}(1980)}]{BYbook}
\bibinfo{author}{\bibfnamefont{J.~P.} \bibnamefont{Boon}} \bibnamefont{and}
  \bibinfo{author}{\bibfnamefont{S.}~\bibnamefont{Yip}},
  \emph{\bibinfo{title}{Molecular Hydrodynamics}}
  (\bibinfo{publisher}{McGraw-Hill, New York}, \bibinfo{year}{1980}).

\bibitem[{\citenamefont{Balucani and Zoppi}(1994)}]{BZbook}
\bibinfo{author}{\bibfnamefont{U.}~\bibnamefont{Balucani}} \bibnamefont{and}
  \bibinfo{author}{\bibfnamefont{M.}~\bibnamefont{Zoppi}},
  \emph{\bibinfo{title}{Dynamics of the Liquid State}}
  (\bibinfo{publisher}{Clarendon Press, Oxford}, \bibinfo{year}{1994}),
  \bibinfo{edition}{2nd} ed.

\bibitem[{\citenamefont{Bengtzelius et~al.}(1984)\citenamefont{Bengtzelius,
  G{\"o}tze, and Sj{\"o}lander}}]{MB}
\bibinfo{author}{\bibfnamefont{U.}~\bibnamefont{Bengtzelius}},
  \bibinfo{author}{\bibfnamefont{W.}~\bibnamefont{G{\"o}tze}},
  \bibnamefont{and}
  \bibinfo{author}{\bibfnamefont{A.}~\bibnamefont{Sj{\"o}lander}},
  \bibinfo{journal}{J.\ Phys.\ C:\ Solid State Physics}
  \textbf{\bibinfo{volume}{17}}, \bibinfo{pages}{5915} (\bibinfo{year}{1984}).

\bibitem[{\citenamefont{Leutheusser}(1984)}]{MC}
\bibinfo{author}{\bibfnamefont{E.}~\bibnamefont{Leutheusser}},
  \bibinfo{journal}{Phys.\ Rev.\ A} \textbf{\bibinfo{volume}{29}},
  \bibinfo{pages}{2765} (\bibinfo{year}{1984}).

\bibitem[{\citenamefont{Kob and Andersen}(1995{\natexlab{a}})}]{LB}
\bibinfo{author}{\bibfnamefont{W.}~\bibnamefont{Kob}} \bibnamefont{and}
  \bibinfo{author}{\bibfnamefont{H.~C.} \bibnamefont{Andersen}},
  \bibinfo{journal}{Phys.\ Rev.\ E} \textbf{\bibinfo{volume}{51}},
  \bibinfo{pages}{4626} (\bibinfo{year}{1995}{\natexlab{a}}).

\bibitem[{\citenamefont{Kob and Andersen}(1995{\natexlab{b}})}]{LC}
\bibinfo{author}{\bibfnamefont{W.}~\bibnamefont{Kob}} \bibnamefont{and}
  \bibinfo{author}{\bibfnamefont{H.~C.} \bibnamefont{Andersen}},
  \bibinfo{journal}{Phys.\ Rev.\ E} \textbf{\bibinfo{volume}{52}},
  \bibinfo{pages}{4134} (\bibinfo{year}{1995}{\natexlab{b}}).

\bibitem[{\citenamefont{G{\"o}tze}(1999)}]{ME}
\bibinfo{author}{\bibfnamefont{W.}~\bibnamefont{G{\"o}tze}},
  \bibinfo{journal}{J.\ Phys.:\ Condens. Matter} \textbf{\bibinfo{volume}{11}},
  \bibinfo{pages}{A1} (\bibinfo{year}{1999}).

\bibitem[{\citenamefont{Copley and Rowe}(1974)}]{CRPRL74}
\bibinfo{author}{\bibfnamefont{J.~R.~D.} \bibnamefont{Copley}}
  \bibnamefont{and} \bibinfo{author}{\bibfnamefont{J.~M.} \bibnamefont{Rowe}},
  \bibinfo{journal}{Phys.\ Rev.\ Lett.} \textbf{\bibinfo{volume}{32}},
  \bibinfo{pages}{49} (\bibinfo{year}{1974}).

\bibitem[{\citenamefont{Rahman}(1974{\natexlab{a}})}]{ARPRL74}
\bibinfo{author}{\bibfnamefont{A.}~\bibnamefont{Rahman}},
  \bibinfo{journal}{Phys.\ Rev.\ Lett.} \textbf{\bibinfo{volume}{32}},
  \bibinfo{pages}{52} (\bibinfo{year}{1974}{\natexlab{a}}).

\bibitem[{\citenamefont{Rahman}(1974{\natexlab{b}})}]{ARPRA74}
\bibinfo{author}{\bibfnamefont{A.}~\bibnamefont{Rahman}},
  \bibinfo{journal}{Phys.\ Rev.\ A} \textbf{\bibinfo{volume}{9}},
  \bibinfo{pages}{1667} (\bibinfo{year}{1974}{\natexlab{b}}).

\bibitem[{\citenamefont{Copley and Lovesey}(1975)}]{CLRPP38}
\bibinfo{author}{\bibfnamefont{J.~R.~D.} \bibnamefont{Copley}}
  \bibnamefont{and} \bibinfo{author}{\bibfnamefont{S.~W.}
  \bibnamefont{Lovesey}}, \bibinfo{journal}{Rep.\ Prog.\ Phys.}
  \textbf{\bibinfo{volume}{38}}, \bibinfo{pages}{461} (\bibinfo{year}{1975}).

\bibitem[{\citenamefont{Bosse et~al.}(1978{\natexlab{a}})\citenamefont{Bosse,
  G{\"o}tze, and L{\"u}cke}}]{BGLPRA17a}
\bibinfo{author}{\bibfnamefont{J.}~\bibnamefont{Bosse}},
  \bibinfo{author}{\bibfnamefont{W.}~\bibnamefont{G{\"o}tze}},
  \bibnamefont{and}
  \bibinfo{author}{\bibfnamefont{M.}~\bibnamefont{L{\"u}cke}},
  \bibinfo{journal}{Phys.\ Rev.\ A} \textbf{\bibinfo{volume}{17}},
  \bibinfo{pages}{434} (\bibinfo{year}{1978}{\natexlab{a}}).

\bibitem[{\citenamefont{Bosse et~al.}(1978{\natexlab{b}})\citenamefont{Bosse,
  G{\"o}tze, and L{\"u}cke}}]{BGLPRA17b}
\bibinfo{author}{\bibfnamefont{J.}~\bibnamefont{Bosse}},
  \bibinfo{author}{\bibfnamefont{W.}~\bibnamefont{G{\"o}tze}},
  \bibnamefont{and}
  \bibinfo{author}{\bibfnamefont{M.}~\bibnamefont{L{\"u}cke}},
  \bibinfo{journal}{Phys.\ Rev.\ A} \textbf{\bibinfo{volume}{17}},
  \bibinfo{pages}{447} (\bibinfo{year}{1978}{\natexlab{b}}).

\bibitem[{\citenamefont{Sj{\"o}gren}(1980{\natexlab{a}})}]{Sjog80}
\bibinfo{author}{\bibfnamefont{L.}~\bibnamefont{Sj{\"o}gren}},
  \bibinfo{journal}{Phys.\ Rev.\ A} \textbf{\bibinfo{volume}{22}},
  \bibinfo{pages}{2866} (\bibinfo{year}{1980}{\natexlab{a}}).

\bibitem[{\citenamefont{Sj{\"o}gren}(1980{\natexlab{b}})}]{Sjog80b}
\bibinfo{author}{\bibfnamefont{L.}~\bibnamefont{Sj{\"o}gren}},
  \bibinfo{journal}{Phys.\ Rev.\ A} \textbf{\bibinfo{volume}{22}},
  \bibinfo{pages}{2883} (\bibinfo{year}{1980}{\natexlab{b}}).

\bibitem[{\citenamefont{Scopigno
  et~al.}(2000{\natexlab{a}})\citenamefont{Scopigno, Balucani, Cunsolo,
  Masciovecchio, Ruocco, Sette, and Verbeni}}]{R&c002}
\bibinfo{author}{\bibfnamefont{T.}~\bibnamefont{Scopigno}},
  \bibinfo{author}{\bibfnamefont{U.}~\bibnamefont{Balucani}},
  \bibinfo{author}{\bibfnamefont{A.}~\bibnamefont{Cunsolo}},
  \bibinfo{author}{\bibfnamefont{C.}~\bibnamefont{Masciovecchio}},
  \bibinfo{author}{\bibfnamefont{G.}~\bibnamefont{Ruocco}},
  \bibinfo{author}{\bibfnamefont{F.}~\bibnamefont{Sette}}, \bibnamefont{and}
  \bibinfo{author}{\bibfnamefont{R.}~\bibnamefont{Verbeni}},
  \bibinfo{journal}{Europhys.\ Lett.} \textbf{\bibinfo{volume}{50}},
  \bibinfo{pages}{189} (\bibinfo{year}{2000}{\natexlab{a}}).

\bibitem[{\citenamefont{Scopigno
  et~al.}(2002{\natexlab{a}})\citenamefont{Scopigno, Balucani, Ruocco, and
  Sette}}]{SBRSb}
\bibinfo{author}{\bibfnamefont{T.}~\bibnamefont{Scopigno}},
  \bibinfo{author}{\bibfnamefont{U.}~\bibnamefont{Balucani}},
  \bibinfo{author}{\bibfnamefont{G.}~\bibnamefont{Ruocco}}, \bibnamefont{and}
  \bibinfo{author}{\bibfnamefont{F.}~\bibnamefont{Sette}},
  \bibinfo{journal}{Phys.\ Rev.\ E} \textbf{\bibinfo{volume}{65}},
  \bibinfo{pages}{031205} (\bibinfo{year}{2002}{\natexlab{a}}).

\bibitem[{\citenamefont{Yulmetyev et~al.}(2003)\citenamefont{Yulmetyev,
  Mokshin, Scopigno, and H{\"a}nggi}}]{Naexp}
\bibinfo{author}{\bibfnamefont{R.~M.} \bibnamefont{Yulmetyev}},
  \bibinfo{author}{\bibfnamefont{A.~V.} \bibnamefont{Mokshin}},
  \bibinfo{author}{\bibfnamefont{T.}~\bibnamefont{Scopigno}}, \bibnamefont{and}
  \bibinfo{author}{\bibfnamefont{P.}~\bibnamefont{H{\"a}nggi}},
  \bibinfo{journal}{J.\ Phys.:\ Condens. Matter} \textbf{\bibinfo{volume}{15}},
  \bibinfo{pages}{2235} (\bibinfo{year}{2003}).

\bibitem[{\citenamefont{Scopigno
  et~al.}(2002{\natexlab{b}})\citenamefont{Scopigno, Ruocco, Sette, and
  Viliani}}]{SRSVPRE02}
\bibinfo{author}{\bibfnamefont{T.}~\bibnamefont{Scopigno}},
  \bibinfo{author}{\bibfnamefont{G.}~\bibnamefont{Ruocco}},
  \bibinfo{author}{\bibfnamefont{F.}~\bibnamefont{Sette}}, \bibnamefont{and}
  \bibinfo{author}{\bibfnamefont{G.}~\bibnamefont{Viliani}},
  \bibinfo{journal}{Phys.\ Rev.\ E} \textbf{\bibinfo{volume}{66}},
  \bibinfo{pages}{031205} (\bibinfo{year}{2002}{\natexlab{b}}).

\bibitem[{\citenamefont{Scopigno
  et~al.}(2002{\natexlab{c}})\citenamefont{Scopigno, Ruocco, Sette, and
  Viliani}}]{SRSVPM02}
\bibinfo{author}{\bibfnamefont{T.}~\bibnamefont{Scopigno}},
  \bibinfo{author}{\bibfnamefont{G.}~\bibnamefont{Ruocco}},
  \bibinfo{author}{\bibfnamefont{F.}~\bibnamefont{Sette}}, \bibnamefont{and}
  \bibinfo{author}{\bibfnamefont{G.}~\bibnamefont{Viliani}},
  \bibinfo{journal}{Phil.\ Mag.\ B} \textbf{\bibinfo{volume}{82}},
  \bibinfo{pages}{233} (\bibinfo{year}{2002}{\natexlab{c}}).

\bibitem[{\citenamefont{Scopigno
  et~al.}(2000{\natexlab{b}})\citenamefont{Scopigno, Balucani, Ruocco, and
  Sette}}]{SBRS}
\bibinfo{author}{\bibfnamefont{T.}~\bibnamefont{Scopigno}},
  \bibinfo{author}{\bibfnamefont{U.}~\bibnamefont{Balucani}},
  \bibinfo{author}{\bibfnamefont{G.}~\bibnamefont{Ruocco}}, \bibnamefont{and}
  \bibinfo{author}{\bibfnamefont{F.}~\bibnamefont{Sette}},
  \bibinfo{journal}{Phys.\ Rev.\ Lett.} \textbf{\bibinfo{volume}{85}},
  \bibinfo{pages}{4076} (\bibinfo{year}{2000}{\natexlab{b}}).

\bibitem[{\citenamefont{Scopigno
  et~al.}(2000{\natexlab{c}})\citenamefont{Scopigno, Balucani, Ruocco, and
  Sette}}]{SBRSa}
\bibinfo{author}{\bibfnamefont{T.}~\bibnamefont{Scopigno}},
  \bibinfo{author}{\bibfnamefont{U.}~\bibnamefont{Balucani}},
  \bibinfo{author}{\bibfnamefont{G.}~\bibnamefont{Ruocco}}, \bibnamefont{and}
  \bibinfo{author}{\bibfnamefont{F.}~\bibnamefont{Sette}},
  \bibinfo{journal}{J.\ Phys.:\ Condens. Matter} \textbf{\bibinfo{volume}{12}},
  \bibinfo{pages}{8009} (\bibinfo{year}{2000}{\natexlab{c}}).

\bibitem[{\citenamefont{Mezei}(1989)}]{Mez89}
\bibinfo{author}{\bibfnamefont{F.}~\bibnamefont{Mezei}},
  \bibinfo{journal}{Springer Proceedings in Physics}
  \textbf{\bibinfo{volume}{37}}, \bibinfo{pages}{164} (\bibinfo{year}{1989}).

\bibitem[{\citenamefont{Brakkee and de~Leeuw}(1989)}]{deLeew88}
\bibinfo{author}{\bibfnamefont{M.~J.~D.} \bibnamefont{Brakkee}}
  \bibnamefont{and} \bibinfo{author}{\bibfnamefont{S.~W.}
  \bibnamefont{de~Leeuw}}, \bibinfo{journal}{Springer Proceedings in Physics}
  \textbf{\bibinfo{volume}{40}}, \bibinfo{pages}{154} (\bibinfo{year}{1989}).

\bibitem[{\citenamefont{Brakkee and de~Leeuw}(1990)}]{deLeew90}
\bibinfo{author}{\bibfnamefont{M.~J.~D.} \bibnamefont{Brakkee}}
  \bibnamefont{and} \bibinfo{author}{\bibfnamefont{S.~W.}
  \bibnamefont{de~Leeuw}}, \bibinfo{journal}{J.\ Phys.:\ Condens. Matter}
  \textbf{\bibinfo{volume}{2}}, \bibinfo{pages}{4991} (\bibinfo{year}{1990}).

\bibitem[{\citenamefont{Mazzacurati et~al.}(1996)\citenamefont{Mazzacurati,
  Ruocco, and Sampoli}}]{S1}
\bibinfo{author}{\bibfnamefont{V.}~\bibnamefont{Mazzacurati}},
  \bibinfo{author}{\bibfnamefont{G.}~\bibnamefont{Ruocco}}, \bibnamefont{and}
  \bibinfo{author}{\bibfnamefont{M.}~\bibnamefont{Sampoli}},
  \bibinfo{journal}{Europhys.\ Lett.} \textbf{\bibinfo{volume}{34}},
  \bibinfo{pages}{681} (\bibinfo{year}{1996}).

\bibitem[{\citenamefont{Ruocco et~al.}(2000)\citenamefont{Ruocco, Sette,
  {Di~Leonardo}, Monaco, Sampoli, Scopigno, and Viliani}}]{R&cPRL00}
\bibinfo{author}{\bibfnamefont{G.}~\bibnamefont{Ruocco}},
  \bibinfo{author}{\bibfnamefont{F.}~\bibnamefont{Sette}},
  \bibinfo{author}{\bibfnamefont{R.}~\bibnamefont{{Di~Leonardo}}},
  \bibinfo{author}{\bibfnamefont{G.}~\bibnamefont{Monaco}},
  \bibinfo{author}{\bibfnamefont{M.}~\bibnamefont{Sampoli}},
  \bibinfo{author}{\bibfnamefont{T.}~\bibnamefont{Scopigno}}, \bibnamefont{and}
  \bibinfo{author}{\bibfnamefont{G.}~\bibnamefont{Viliani}},
  \bibinfo{journal}{Phys.\ Rev.\ Lett.} \textbf{\bibinfo{volume}{84}},
  \bibinfo{pages}{5788} (\bibinfo{year}{2000}).

\end{thebibliography}

\end{document}